\documentclass[superscriptaddress,floatfix,amsmath,amssymb,aps,prx,reprint]{revtex4-2}
\usepackage{physics}
\usepackage{bbold}
\usepackage{natbib}    
\usepackage{color}         
\usepackage{graphicx}      
\usepackage{epsfig}          
\usepackage{bm}            
\usepackage{amsmath}
\usepackage{thumbpdf}
\usepackage{hyperref} 
\usepackage{placeins}
\usepackage{xcolor}
\usepackage{multirow}
\usepackage{caption}
\usepackage[normalem]{ulem}

\begin{document}

\title{Deterministic domain selection of antiferromagnets via magnetic fields}
\author{Sophie F. Weber}
\affiliation{Department of Physics and Astronomy, Chalmers University of Technology, SE-412 96 Gothenburg, Sweden}
\author{Veronika Sunko}
\affiliation{Institute of Science and Technology Austria (ISTA), 3400 Klosterneuburg, Austria} 
\date{\today}
\begin{abstract}
Antiferromagnets (AFMs) hold promise for applications in digital logic. However, switching AFM domains is challenging, as magnetic fields do not couple to the bulk antiferromagnetic order parameter. Here we show that magnetic-field-driven switching of AFM domains can in many cases be enabled by a generic reduction of magnetic exchange at surfaces. We use statistical mechanics and Monte Carlo simulations to demonstrate that an inequivalence in magnetic exchange between top and bottom surface moments, combined with the enhanced magnetic susceptibility of surface spins, can enable deterministic selection of a single AFM domain, with the selected domain depending on the magnetic field ramping direction. We further show that this mechanism naturally explains an observed hysteretic magneto-optical response of even-layer van der Waals AFM $\mathrm{MnBi_2Te_4}$. Our findings highlight the critical role of surface spins in the magnetic field responses of AFMs, and reveal a previously unrecognized route to field-based AFM control that requires no net uncompensated moment.
\end{abstract}

\maketitle

Antiferromagnets (AFMs) are promising spintronics materials because their staggered magnetization, or N\'{e}el vector, is a bit-like quantity suitable for logic and storage. They also offer advantages over ferromagnets (FMs) such as robustness against stray magnetic fields and ultrafast dynamics~\cite{Olejnk2018,Baltz2018}.\\ 
\begin{figure}
    \centering
   \includegraphics[width=0.8\linewidth]{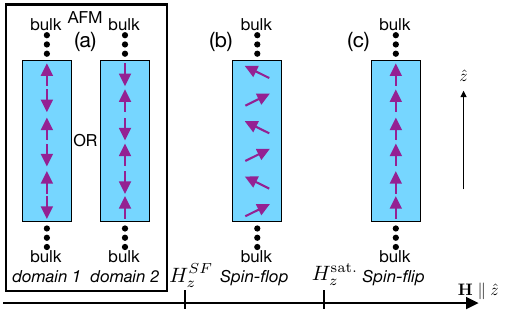}
    \caption{Field response of a periodic bulk A-type AFM as a function of magnetic field strength $H_z$ applied parallel to the easy axis. $H_z^{SF}$ and $H_z^{\mathrm{sat.}}$ denote the spin flop and the saturation field, respectively. }
    \label{fig:bfield_dependence}
\end{figure}
\indent While the insensitivity of AFMs to moderate magnetic fields is beneficial for stability, it is thought to hinder deterministic control and switching of the N\'{e}el vector. For fields weak compared to the AFM Heisenberg coupling $J^{\mathrm{AFM}}$, the two AFM domains are degenerate (Fig.~\ref{fig:bfield_dependence}(a)). Stronger fields drive spin-flop (Fig.~\ref{fig:bfield_dependence}(b)) and spin-flip (Fig.~\ref{fig:bfield_dependence}(c)) transitions respectively, altering AFM order rather than enabling domain selection.\\ 
\indent On the other hand, deterministic selection of AFM domains by a magnetic field has been reported in the layered antiferromagnet $\mathrm{MnBi_2Te_4}$ (MBT). This is reflected in observed electrodynamic properties such as a magneto-optical Kerr effect (MOKE) whose sign is determined by the N\'{e}el vector~\cite{ovchinnikov_intertwined_2021, yang_odd-even_2021, Qiu2023, bartram_real-time_2023, Sunko2026}. MBT is an A-type AFM structure, with out-of-plane spins in each layer parallel to each other, and anti-parallel to those in neighboring layers. While reported domain selection for odd-layer, ferrimagnetic MBT has been explained via field-pinning of uncompensated magnetization~\cite{Wang2025,Chong2024}, the experiments of interest here report domain selection in even-layer, zero-net-magnetization MBT~\cite{Qiu2023, yang_odd-even_2021,  ovchinnikov_intertwined_2021, Sunko2026}. Thus, based on Fig.~\ref{fig:bfield_dependence}, field-based domain selection in these cases should be impossible.\\
\indent In this work, we resolve this apparent contradiction by considering a key feature of real materials; surfaces, which are known to exhibit behavior distinct from the corresponding bulk~\cite{Belashchenko2010,Sunko2017, Spaldin2021,Bhowal2025,Weber2023,Weber2024,Seleznev2023}.
 Using statistical mechanics and Monte Carlo (MC) simulations, we demonstrate that reduced effective Heisenberg exchange at the surfaces of AFMs with no net magnetization can drive the magnetic field response of the entire slab. In particular, if the AFM exchange on top and bottom surfaces is sufficiently different, ramping $\left|H_z\right|$ from above $\left|H_z^{SF}\right|$ to $H_z=0$ can lead to \textit{deterministic} selection of a single AFM domain, with the N\'{e}el vector sign dependent on the direction of ramping. 
 We use our MC results to model the MOKE signal as a function of magnetic field for thick flakes of MBT, and find very good agreement with recent experiments~\cite{Sunko2026}.\\
\indent Although MBT inspired this work, our results apply broadly to AFMs with magnetically uncompensated surfaces (while maintaining zero net magnetization in the full AFM) and an AFM propagation-vector component perpendicular to the surface. Given that inequivalence between opposite crystal surfaces is ubiquitous (for example, via growth on a substrate, interfacing with another material, or defects), magnetic field-based AFM domain selection is likely to be realized in many experimental systems. \\
\section{\label{sec:statmech}Statistical Mechanics considerations} 
\begin{figure}
    \centering
   \includegraphics[width=\columnwidth]{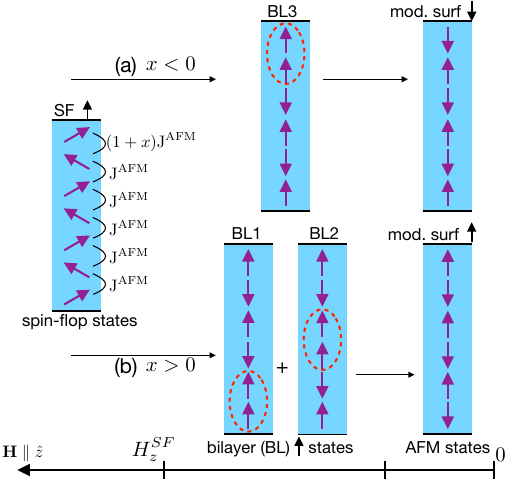}
    \caption{Transition with decreasing applied field for an asymmetric AFM slab from spin-flop state (left) to one or more intermediate ``bilayer" phases (center) to the most probable collinear AFM domain (right) as $H_z\rightarrow0$, reflecting a ramping-down protocol.  Layers 1-5 have interlayer couplings $J^{\mathrm{AFM}}$, while the top coupling is $(1+x)J^{\mathrm{AFM}}$. (a) Transition pathway for $x<0$. (b) Pathway for $x>0$.}
    \label{fig:bl_cartoon}
\end{figure}

As a minimal model we consider a layered AFM with one spin per layer in the unit cell (we assume uniform intralayer FM coupling), interlayer Heisenberg exchange $J^{\mathrm{AFM}}$, and uniaxial single-ion anisotropy $K$ along $\hat{z}$, with an applied out-of-plane field $H_z$. Analogous arguments hold when the AFM is easy-plane, and $H$ is applied parallel to the layers. The Hamiltonian is given by 
\begin{equation}
    E=E_0+J^{\mathrm{AFM}}\sum_{ij}\hat{e}_i\cdot\hat{e}_j-K\sum_i(\hat{e}_i\cdot\hat{z})^2-C\sum_i(S_i)^zH_z.
    \label{eq:spinH}
\end{equation}
Here, the sum is over magnetic atoms in the unit cell; $\hat{e}_i$ is the unit vector parallel to the magnetic moment on site $i$; $S_{i}$ is the magnetic moment of site $i$ in $\mu_B$, and $C$ converts $\mu_B\cdot\mathrm{T}$ to eV. We start with a $+\hat{z}$-oriented $H_z$ above the spin-flop field $H_z^{\mathrm{SF}}$ (Fig.~\ref{fig:bl_cartoon}(a)). We then consider how the presence of vacuum-terminated surfaces can alter the evolution to a collinear AFM domain as $+H_z\rightarrow 0\rightarrow-H_z$ (a ``ramp-down" procedure).\\
\indent In a bulk AFM, each layer has two neighboring layers, yielding an effective exchange $J^{\mathrm{tot}}=2J^{\mathrm{AFM}}$. In contrast, surface spins have one neighboring layer, and a \textit{halved} total exchange compared to interior spins. As a result, the AFM coupling of surface spins overcomes the Zeeman energy at a lower field than interior spins, stabilizing a field window below $H_z^{SF}$ where both surface spins align with $H_z$. This enforces a multidomain AFM state. In our toy model the domain wall corresponds to a ferromagnetically aligned bilayer (BL); the center panel of Figs.~\ref{fig:bl_cartoon}(a,b) shows its three possible positions for a slab with $N=6$ layers, labeling the corresponding configurations BL1-3. Note that for the BL phase to be stabilized via field pinning of the surface spins, the surface must be magnetically uncompensated. While this is satisfied naturally for the cleavage plane of A-type AFMs, it is also possible for certain terminations of other AFM structures~\cite{Weber2024}.\\ 
\indent Reduced magnetic exchange is not enough to enable deterministic domain selection: the exchanges on top and bottom surfaces also need to be inequivalent. In the following, we assume that all interlayer exchanges maintain the bulk $J^{\mathrm{AFM}}$ except for coupling at one of the two surfaces, which varies as $(1+x)J^{\mathrm{AFM}}$ with $-1<x<1$ (Fig.~\ref{fig:bl_cartoon}, left). In the cartoon, the inequivalent exchange is at the top surface, but analogous arguments hold when the modified exchange is at the bottom surface, as well as when \textit{both} surface exchanges are inequivalently modified from bulk values (SI appendix, Sec.~H). Now, a state with the BL formed at the surface with modified exchange is lower ($x<0$) or higher ($x>0$) in energy than all other BL states, and is thus more or less likely (See SI appendix Sec.~K for energetics of spin-flop, AFM, and BL states as function of applied field).\\ 
\indent For $x<0$ (Fig.~\ref{fig:bl_cartoon}(a)), BL3 dominates at fields just below $|H_z^{SF}|$ when ramping down (center panel). As $|H_z|$ is further reduced, the AFM domain whose spin at the surface with modified exchange is \emph{opposite} the field direction (mod. surf $\downarrow$) will be favored, since it differs from BL3 by a single spin flip, whereas reaching the other AFM domain would require flipping all but the top spin. The opposite domain, mod. surf $\uparrow$, is favored for a ramp-up protocol. For $x>0$ (Fig.~\ref{fig:bl_cartoon}(b)), using similar arguments we find that opposite domains are more likely for a given ramping protocol compared to $x<0$. In summary, at $H_z =
0$, the AFM domain in which the spin on the surface with
weaker AFM coupling points opposite the initial field direction
is favored. We quantify the above arguments in SI appendix, Secs.~F and~G via committor analysis.\\
\indent We mention that the first stage in deterministic domain selection (center panel, Fig.~\ref{fig:bl_cartoon}) is based on selection of lowest-energy BL states. In this sense, it has close connections to exchange bias, where interfacial coupling sets an energetically preferred AFM domain in an AFM-FM heterostructure~\cite{Nogues1999}. However, inequivalent surface exchange \textit{does not} lift the energetic degeneracy of the two zero-magnetization AFM monodomains as the field is further lowered. Rather, in the second step (Fig.~\ref{fig:bl_cartoon}, right panel) the selected BL position breaks degeneracy in subsequent \textit{pathway probability} for the two monodomains. In SI Sec.~M, we use Monte Carlo to contrast our proposed two-step mechanism relevant for zero-magnetization, six-layer MBT with purely exchange bias-based selection involving uncompensated magnetization in \textit{odd-layer}, ferrimagnetic MBT.\\
\indent It is also crucial to note that the $x>0$ and $x<0$ scenarios are not equivalent, and in particular have different scaling with respect to number of layers. For $x<0$, the state with the BL localized at the weakened surface, which leads to subsequent deterministic monodomain selection, is lower in energy than the $\frac{N}{2}-1$ states with other BL positions by $\Delta E=2N_{\parallel}J^{\mathrm{AFM}}|x|$, assuming $N_{\parallel}$ ferromagnetically coupled spins per layer. Defining $\epsilon=P({\rm BL,other})/P({\rm BL^\ast})$ as the probability for entering any other BL state versus this lower-energy surface-localized BL state, the critical surface exchange asymmetry $|x^*|$ for deterministic domain selection, given a maximum allowed $\epsilon^*$, satisfies
\begin{equation}
|x^*|\sim
\frac{k_BT}{2N_{\rm coh}J^{\rm AFM}}
\ln\left(\frac{N}{2\epsilon^\ast}\right).
\label{eq:xval}
\end{equation}
where $N_{\rm coh}<N_{\parallel}$ is the number of in-plane spins that coherently initiate the transition to a specific BL state.
The logarithmic dependence on $N$ (see SI appendix Sec.~E) implies that the necessary surface exchange anisotropy remains experimentally plausible up to bulk-like thicknesses.\\
\indent For $x>0$, by contrast, the state with the BL at the modified surface is energetically disfavored, while the opposite-surface BL state, which selects the opposite AFM domain, competes with $O(N)$ energetically degenerate central BL positions. Assuming equal accessibility from the spin-flop phase to all degenerate BL states, this gives a finite domain bias, rather than fully deterministic selection for $x>0$, even for thin films, with the relevant surface-BL weight scaling as $\sim 1/N$. Interestingly, our Monte Carlo simulations for thin-film, six-layer MBT do show deterministic selection for both signs of $x$, suggesting that pathways to surface-localized BL states can be kinetically enhanced enough to produce deterministic selection when the number of energetically competing central BL states is small (SI appendix Sec.~J). In the following, we focus on the $x<0$ case relevant for bulk-like thicknesses.
\section{\label{sec:MC}Monte Carlo simulations}
To test whether our proposed mechanism can cause the deterministic selection in thick samples implied by the MOKE measurements of MBT in Ref~\cite{Sunko2026}, we perform Monte Carlo (MC) simulations of MBT with in-plane periodicity and out-of-plane vacuum boundary conditions. We use an in-plane ferromagnetic coupling, uniaxial anisotropy,  and out-of-plane $J^{\mathrm{AFM}}$ based on the density functional theory (DFT) values obtained for MBT in Refs.~\cite
{Otrokov2019,Otrokov2019B} (See SI appendix Sec.~A for details).  We model a thick slab of 48 layers ($\sim64~\mathrm{nm}$, about half the experimental thickness~\cite{Sunko2026}). We then introduce varying exchange anisotropy $x$ and simulate field ramping at $T=15\mathrm{K}$.\\
\begin{figure*}
    \centering
   \includegraphics{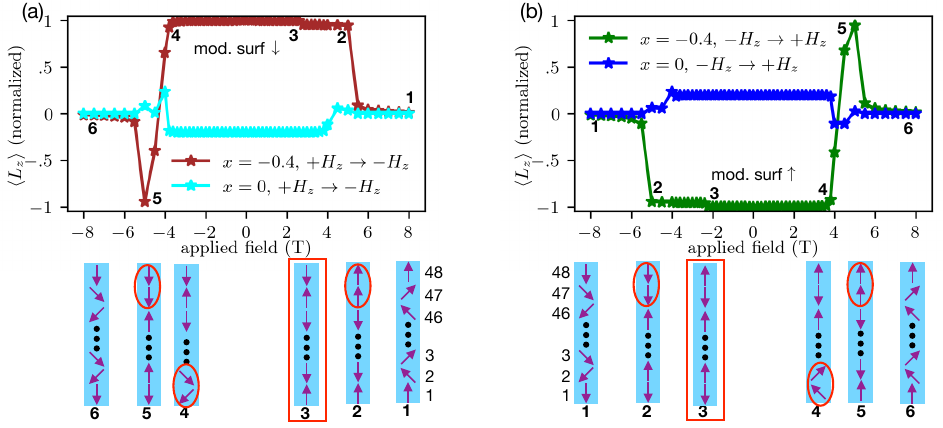}
    \caption{(a) Ensemble-averaged Néel vector $\langle L_z\rangle$ versus applied field from MC simulations of $48$-layer MBT with $x=0$ and $x=-0.4$. Results are shown for (a) ramping-down and (b) ramping-up protocols. Cartoons indicate layer-resolved magnetization at selected fields labeled $1$-$6$ for the top- and bottom-most layers for $x=-0.4$; boxes mark the selected AFM domain, and red circles mark FM-coupled bilayers.
}
    \label{fig:magstates}
\end{figure*}
\indent Figures~\ref{fig:magstates}(a) and (b) show our results for ramping-down and ramping-up protocols respectively. We calculate as a function of field $H_z$ the normalized $\hat{z}$-component of the Néel vector,
\begin{equation}
L_z=\frac{1}{N|m|}\sum_{k=1}^N m_z^k(-1)^{k-1},
\label{eq:Neelvec}
\end{equation}
where $m_z^k$ is the planar-averaged Mn moment in layer $k$ and $|m|=3.9~\mu_B$ is the MC saturation magnetization per Mn at $15\,\mathrm{K}$. In Fig. \ref{fig:magstates} we plot an ensemble average $\langle L_z\rangle$ over five separate MC field-ramping simulations. $\langle L_z \rangle=+1$ ($\langle L_z \rangle=-1$) means that that the mod surf. $\downarrow$ ($\uparrow$) domain was selected in all five simulations.\\
\indent For a symmetric slab ($x=0$), the selected domain in Fig.~\ref{fig:magstates} is random ($\langle L_z\rangle \sim0$). For an asymmetric slab with $x=-0.4$ however, the direction of ramping deterministically selects the expected AFM domain (results for other $x$ are shown in SI appendix Sec.~I). Generalizing beyond MBT, we verify in SI appendix Sec.~C that deterministic domain selection remains robust over a broad range of uniaxial anisotropies, only breaking down when the anisotropy becomes too weak to stabilize the intermediate BL regime. We also show in SI appendix Sec.~D that deterministic selection is not sensitive to ramping rate.\\
\indent We now describe in more detail the magnetic state evolution for the ramp-down protocol, Fig.~\ref{fig:magstates}(a); the ramp-up case is just mirrored in sign. From $H=8~\mathrm{T}$ to $H=\sim5.25~\mathrm{T}$, MBT is in the spin-flop phase with spins canted along $+\hat{z}$ (cartoon 1, Fig.~\ref{fig:magstates}(a)). Due to their enhanced susceptibility, the top and bottom layer spins are already almost fully parallel with $H_z$. At $H_z^{SF}\sim5.25~\mathrm{T}$ the slab transitions to a BL $\uparrow$ state with the FM-coupled bilayer at the surface with weaker exchange (cartoon 2), and $\langle L_z\rangle$ jumps to slightly below $+1$. This is followed by a transition to the preferred AFM monodomain (mod. surf $\downarrow$) at $+2.6~\mathrm{T}$ (cartoon 3), at which point $\langle L_z\rangle$ saturates at $+1$. The monodomain is stable until $-3.8~\mathrm{T}$, when the spins on the surface with stronger exchange begin rotating to align with the negative field (cartoon 4) leading to FM alignment with the layer just above. As the field is further decreased the BL (or domain wall) propagates through the slab until the energetically favorable state with the BL localized at the opposite surface forms around  $-5\mathrm{T}$ (cartoon 5). Finally, at $\sim6\mathrm{T}$ MBT transitions to the spin-flop state with spins canted down, and $\langle L_z\rangle$ vanishes (cartoon 6).
\begin{figure}
    \centering\vspace{-.5cm}
   \includegraphics{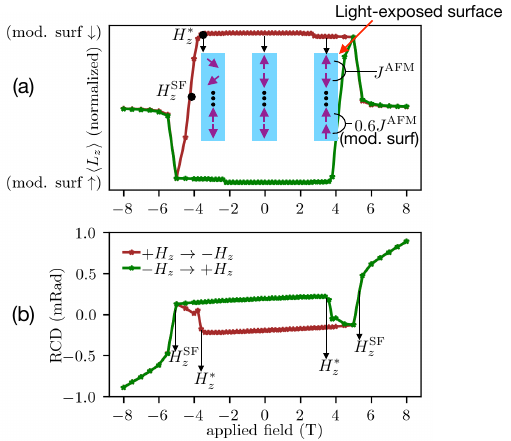}
    \caption{(a) Repitition of Fig.~\ref{fig:magstates} for $x=-0.4$ with ramping-up and ramping-down superimposed. Cartoons show how in the RCD simulation (b), the surface with  weaker exchange is at the bottom of the sample, invisible to the light. (b) Simulated RCD measurement using the MC data in (a). The simulation is done for the photon energy of $2.3\text{eV}$, using the same dielectric model as in Ref.~\cite{Sunko2026}.}
    \label{fig:MC_LvsH}
\end{figure}
\section{\label{sec:MOKE}Comparison to MOKE on MnBi$_2$Te$_4$} We now follow the procedure in Ref.~\cite{Sunko2026} to simulate the MOKE signal arising from the magnetic structures in our MC simulations (see SI appendix, Sec.~L for details).\\
\indent In Fig. \ref{fig:MC_LvsH}(b) we show the simulated reflection circular dichroism (RCD, i.e. imaginary part of MOKE) data based on the MC layer-resolved magnetism for the deterministic case of $x=-0.4$ described above. Fig.~\ref{fig:magstates} is plotted again in Fig. \ref{fig:MC_LvsH}(a) for ease of comparison to the MOKE simulation, with ramping-down (brown) and ramping-up (green) superimposed. The RCD simulation is done assuming that the surface with stronger exchange is exposed to the light since this yields the best agreement with experiment~\cite{Sunko2026}. The agreement with the experimental RCD (Fig.~2 in Ref.~\cite{Sunko2026}) is quite good. The main qualitative feature, a hysteretic RCD signal, is reproduced. Moving beyond that, we observe two characteristic fields, $H_z^{SF}=\pm5.25~\mathrm{T}$ and $H_z^*=\pm3.8~\mathrm{T}$, which is also a feature of the experiment (although the field values differ). The MC simulations, described above, clarify the origin of the two fields: $H_z^{SF}$ and $H_z^*$ correspond to transitioning between a spin-flop to multidomain BL state (Fig.~\ref{fig:magstates}(a) cartoons 1-2), and from an AFM monodomain to a multidomain state (cartoons 3-5), respectively. Note that the other characteristic field visible in $\langle L_z\rangle$, the BL to modomain AFM transition at $H_z\sim\pm2.6\mathrm{T}$, is not evident in Fig.~\ref{fig:magstates}(b). This is because in the RCD simulation, the lowest-energy state in the BL phase has the bilayer located at the weaker-exchange surface, which is invisible to the light. Thus, the single layer flip at this surface in transitioning to the monodomain is not visible. As we see in SI appendix, Sec.~L, if the RCD spectrum is instead modeled assuming the surface with weakened exchange is exposed to light, both $H_z^*$ and this additional characteristic field are evident.\\
\indent Our findings in Fig.~\ref{fig:MC_LvsH} also reveal why $H_z^{SF}$ is highly reproducible, while $H^{*}$ is found to vary between experimental studies, and even among different positions on the same sample~\cite
{Sunko2026}. $H_z^{SF}$ is set by bulk quantities (AFM exchange and Zeeman energy). In contrast, $H_z^{*}$ is set by surface properties, and corresponds to a reconfiguration of AFM domains, which may be strongly affected by details induced by distinct sample preparation protocols.\\
\indent Finally, to assess whether the surface exchange asymmetry required for deterministic domain selection in our MC simulations is physically realistic, we performed density functional theory (DFT) calculations for two representative mechanisms that naturally make the two MBT surfaces inequivalent: oxidation of an air-exposed surface~\cite{mazza_surface-driven_2022} and growth on a substrate, where we chose $(111)$-oriented $\mathrm{BaF_2}$~\cite{Kagerer2020} (see SI appendix Sec.~B). For the modified surfaces we obtain $x$ values in the ranging from $-0.2$ (for full-coverage oxidation) to $+0.2$ (for substrate growth). These values are more than sufficient for deterministic selection in thin films (SI appendix Sec.~J), and are within a factor of two of the $x=-0.4$ value needed in the bulk-like slab in Figs.~\ref{fig:magstates} and ~\ref{fig:MC_LvsH}. Given the approximations inherent in our MC simulations, and the fact that we have sampled only a small subset of possible experimental situations in DFT, this agreement is quite good and supports the experimental plausibility of the phenomenological asymmetries considered here.\\
\section{Conclusions} We first comment on this work's implications for MBT. Our MC simulations show that sufficient asymmetry between opposite surface exchanges produces ramp-direction-dependent domain selection and field-dependent MOKE consistent with experiment. The mechanism also explains why switching fields can vary with sample preparation~\cite{li_fabrication-induced_2024}, coupling to substrates~\cite{li_observation_2022}, or oxidation.\\
\indent Our results further suggest which material platforms may be most promising for realizing deterministic domain selection. Magnetic field-based domain control persists beyond the ultra thin-film limit only when at least one surface exhibits weaker—rather than stronger—interlayer AFM exchange than the bulk. Based on our DFT results, defects or oxidation likely favor this regime, while substrates may either weaken or strengthen interfacial AFM exchange, depending on epitaxial strain, suggesting a design strategy for deterministic control.\\
\indent More broadly, this work reveals a previously unrecognized possibility for magnetic field-based domain control in zero-magnetization AFMs that does not require any interfaced ferromagnetic layer or weak intrinsic ferromagnetic component. The mechanism exploits an exchange bias-like surface-pinning step, but deterministic selection of the final AFM domain arises from subsequent pathway asymmetry rather than from energetic bias between the two AFM monodomains. The required ingredients for the mechanism are the generic reduced magnetic exchange at surfaces, combined with asymmetric exchange of top and bottom surfaces, which occurs in many experimental scenarios.\\ 
 \indent Given this, magnetic field ramping should be a promising, long-overlooked option for domain selection in completely magnetically compensated AFMs. Going forward, controlled experiments and ab-initio calculations need to be conducted to identify the materials and conditions where this mechanism is most promising.

\textbf{Acknowledgments}\\
SFW acknowledges funding from Chalmers University of Technology through the department of Physics and Astronomy, and the Areas of Advance Nano and Materials Science. VS acknowledges funding from Institute of Science and Technology Austria. Monte Carlo simulations were performed using computing resources from the PDC Center for High Performance Computing.~These resources were granted by the National Academic Infrastructure for Supercomputing in Sweden (NAISS), partially funded by the Swedish Research Council through grant agreement no. 2022-06725.\\
\nocite{*}
\FloatBarrier            
\bibliography{AFMdomain_selection.bib}

\clearpage
\counterwithout{equation}{section} \renewcommand\theequation{S\arabic{equation}} \renewcommand\thefigure{S\arabic{figure}} \renewcommand\thetable{S\arabic{table}} \renewcommand\thesection{S\arabic{section}} \renewcommand*{\bibnumfmt}[1]{[S#1]} \renewcommand\bibnumfmt[1]{[S#1]} \setcounter{equation}{0} \setcounter{figure}{0} \setcounter{enumiv}{0} \newpage 

\newpage
\setcounter{page}{1} 

\section*{Supplementary Information}

\subsection{Monte Carlo calculations for $\mathrm{MnBi_2Te_4}$}
\begin{figure}
    \centering
   \includegraphics{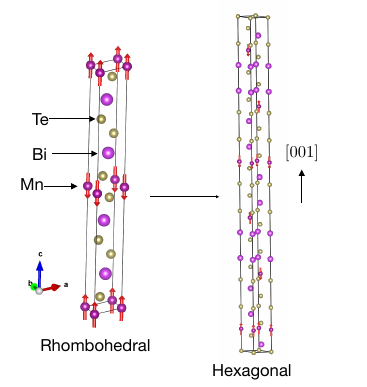}
    \caption{Crystal structure of $\mathrm{MnBi_2Te_4}$ with the ground-state A-type antiferromagnetic ordering in the primitive rhombohedral setting (left) and in the hexagonal setting (right) which we use as the input cell for our Monte Carlo simulations. We model the $(001)$ hexagonal surface cut, reflecting the experimental surface crystal orientation in Ref.~\cite{Sunko2026}, by using vacuum boundary conditions along the $(001)$ direction and periodic in-plane boundary conditions. The picture of the hexagonal structure shows six $\mathrm{Mn}$ layers.}
    \label{fig:MBT}
    \vspace{0.5cm}
    \centering
   \includegraphics{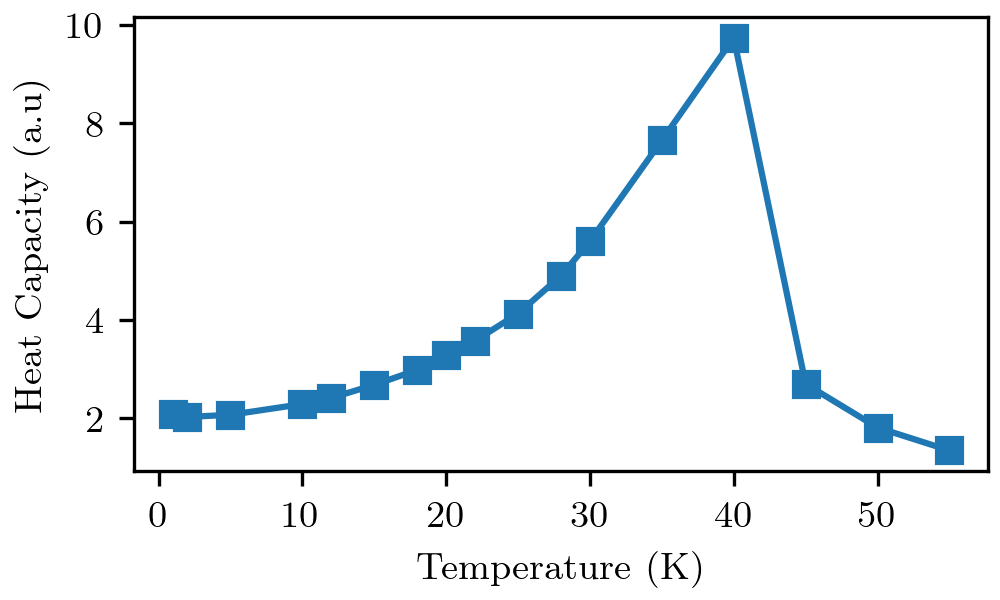}
   \caption{Heat capacity as a function of temperature from a Monte Carlo simulation of bulk $\mathrm{MnBi_2Te_4}$. using the spin Hamiltonian parameters of Ref. ~\cite{Otrokov2019B}.}
   \label{fig:Cv}
\end{figure}

For our simulations of $\mathrm{MnBi_2Te_4}$ (MBT), we use Monte Carlo (MC) as implemented in the spin dynamics package UppASD~\cite{Skubic2008}. We take the hexagonal setting of MBT (Fig. \ref{fig:MBT}) as the input structure, and use supercells of $22\times22\times N$, where $N$ is the number of out-of-plane layers. The correspondingly modest number of in-plane spins ($484$ per layer), is necessary for the 48-layer results shown in the main text due to computational overhead, and we keep the same in-plane size MC simulations of a six-layer MBT slab shown in SI Sec.~J to directly compare with the 48 layer case. We use periodic boundary conditions in the in-plane directions and vacuum boundary conditions along the hexagonal $(001)$ direction to mimic the experimental crystal orientation.\\
\indent We take directly the spin Hamiltonian parameters reported in Refs.~\cite{Otrokov2019} and~\cite{Otrokov2019B} which were obtained from density functional calculations. We include up to third nearest neighbor in-plane Heisenberg exchange, with $J_1=-1.86~\mathrm{meV}$, $J_2=0.05~\mathrm{meV}$ and $J_3=-0.01~\mathrm{meV}$ (where $J<0$ indicates a ferromagnetic coupling and $J>0$ indicates an antiferromagnetic coupling). This yields an effective ferromagnetic in-plane coupling of $J^{\mathrm{FM}}\sim-1.82~\mathrm{meV}$ as mentioned in the main text. For simplicity, for the out-of-plane AFM coupling we use a single effective nearest-neighbor exchange $J^{\mathrm{AFM}}=\sum_{i,\perp}J_i^{\perp}=0.47~\mathrm{meV}$, where the sum is over all out-of-plane nearest neighbors. We use a uniaxial out-of-plane single-ion anisotropy of $0.225~\mathrm{meV}/\mathrm{Mn}$. We take $S_i=4.5~\mu_B$ as the zero-Kelvin value of the $\mathrm{Mn}$ magnetic moments (based on DFT calculations we performed for Ref.~\cite{Sunko2026}). This is the number which is used as the baesline to renormalize the moment magnitudes at finite temperatures in the MC simulations.\\
\indent As an aside, we note that when we calculate the heat capacity using an MC simulation of bulk $\mathrm{MnBi_2Te_4}$ with $24200$ spins (Fig.~\ref{fig:Cv}), we find a N\'{e}el temperature of $T_N\sim 40~\mathrm{K}$ based on the peak. This overestimates the experimental N\'{e}el temperature $T_N\sim25~\mathrm{K}$ of MBT,indicating an overestimation of the experimental exchange values. This is consistent with the fact that the critical field values in our MC simulations are all slightly higher than in experiments Ref.~\cite{Otrokov2019B, Sunko2026}. More surprisingly, our determination of $T_N$ is also in disagreement with the estimate of Ref.~\cite{Otrokov2019B}, from which we obtain our spin Hamiltonian parameters; based on calculation of the susceptibility, they find $T_N\sim25~\mathrm{K}$, very close to experiment. We assume that the discrepancy stems partially from the fact that we compress the out-of-plane coupling into a single effective $J^{\mathrm{AFM}}=\sum_{i,\perp}J_i^{\perp}=0.47~\mathrm{meV}$ for simplicity in making the main point of our work, which is the deterministic selection of AFM domains via magnetic fields when surface exchanges are inequivalent. On the other hand, Ref.~\cite{Otrokov2019B} explicitly takes into account up to 70 magnetic shells of out-of-plane nearest neighbors. Moreover, there are known issues in accurately determining $T_N$ in quasi-2D AFMs such as $\mathrm{MnBi_2Te4}$ due to short-range correlations that can occur well above the true $T_N$~\cite{Hergett2025}. A careful determination of $T_N$ using e.g. Binder cumulants or finite-size scaling in Monte Carlo is beyond the scope and purpose of the present work.\\
\indent For the field-ramping simulations for both 48-layer and six-layer MBT slabs, we use a constant temperature of $15~\mathrm{K}$ and apply an external magnetic field parallel to the out-of-plane easy axis. We start at $8~\mathrm{T}$ ($-8~\mathrm{T}$) for ramping down (ramping up) simulations, Which is above the spin-flop field $|H_z^{SF}|=5.25~\mathrm{T}$ based on our simulations. We initialize a random initial spin configuration for the first field values. After a full MC run at constant field value, we decrease (increase) the field for ramping-down (ramping-up) protocols and run another MC simulation at a new field value, using the final spin configuration from the previous field run as our starting configuration. We ramp the field in steps of $0.5~\mathrm{T}$ between $\pm4~\mathrm{T}$ to $\pm8~\mathrm{T}$ and in steps of $0.2~\mathrm{T}$ between $-4~\mathrm{T}$ and $+4~\mathrm{T}$.
\\
\indent At each field value, the N\'{e}el vector (Eq. 3 in the main text) is calculated using the $\hat{z}$ component of the average sublattice magnetizations $m_z^k$ for each layer $k$, calculated as 
\begin{equation}
m_z^k=\frac{1}{N_{\mathrm{cells}}}\sum_{j=1}^{N_{\mathrm{cells}}}m_{z,j}^k.
\label{eq:mz}
\end{equation}
Here, $N_{\mathrm{cells}}$ is the the number of unit cells (and spins per layer) in the simulation box, $22\times22=484$. We additionally average the layer magnetization in Eq. \ref{eq:mz} over $200,000$ MC steps, starting the average after $500,000$ initial steps which bring the system into equilibrium at each field value. Lastly, we do an ensemble average over five independent MC calculations to get the final values of $m_z^k$ which go into calculation of the field-dependent ensemble-averaged N\'{e}el vector $\langle L_z\rangle$ as well as the MOKE simulations. The relatively few ensemble averages, necessary due to the long run-time of the simulations, result in rough statistics, but are sufficient to make our points.\\
\subsection{DFT calculations}
\begin{figure*}
    \centering
   \includegraphics{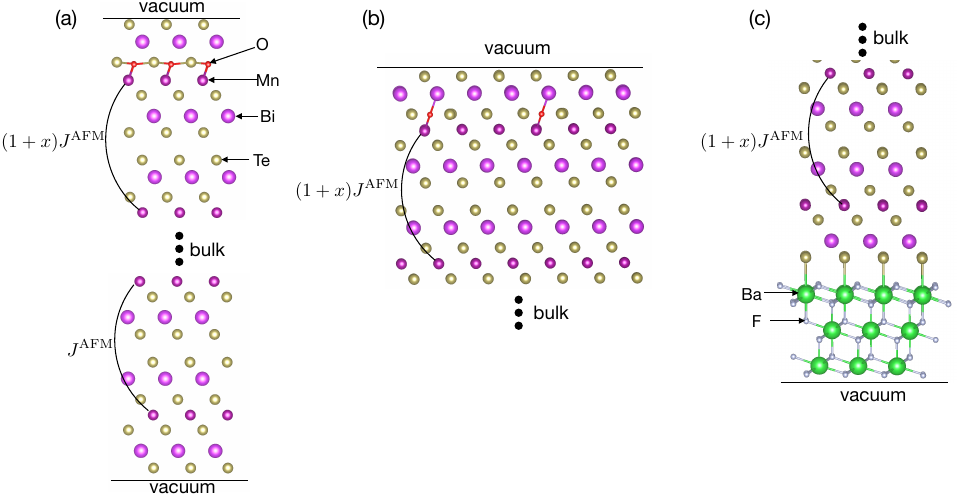}
    \caption{Structures used in our DFT calculations to determine values of exchange anisotropy $x$ in a six-layer, $(001)$-oriented MBT due to inequivalent top and bottom surface environments for a few plausible experimental conditions. (a) Vacuum-terminated MBT with one oxidized surface (the other surface, shown here, is kept pristine). For this simulation, we introduce one oxygen per top surface $\mathrm{Mn}$. (b) Same situation as (a), but with oxidation in the ``dilute" limit, where a $3\times 3\times1$ supercell is used to isolate the oxygen defects. (c) MBT interfaced at one surface with $(111)$-oriented $\mathrm{BaF_2}$, to model growth on a substrate.}
    \label{fig:DFT}
\end{figure*}

As already stated, for all of our Monte Carlo simulations of MBT in the main text and the SI we take ab-initio magnetic parameters from prior litererature~\cite{Otrokov2019,Otrokov2019B}. This is to avoid the computational overhead required to calculate in DFT all necessary spin Hamiltonian inputs, including multiple in-plane ferromagnetic exchanges and unixial anisotropy in addition to the interlayer AFM coupling of focus in this manuscript. However, we do perform a rough set of DFT calculations to gauge whether the order-of-magnitude exchange anisotropies we had to use to achieve fully deterministic domain selection in the MC simulations ($x\sim\pm0.1$ for thin-film, six-layer MBT, see SI Sec.~J, and $x\sim-0.4$ for the bulk-like 48-layer slab discussed in the main text) are plausible.\\
\indent While an exhaustive ab-initio survey of all experimentally relevant sources of surface asymmetry is beyond the scope of the present work, we focus on two realistic and qualitatively distinct mechanisms: oxidation of an air-exposed surface, which is known to occur in MBT~\cite{mazza_surface-driven_2022}; and growth on a substrate. In both cases, we modify just one surface and determine ``$x$'' implied by the value of interlayer $\mathrm{Mn}$ AFM exchange for the modified surface relative to the calculated exchange for the opposite, pristine, vacuum-terminated surface (Fig.~\ref{fig:DFT}(a)). For oxidation simulations, we take two limiting cases; in once case, we introduce one oxygen atom per surface $\mathrm{Mn}$, as in Fig.~\ref{fig:DFT}(a). In the second limit of oxidation, we consider dilute oxygen defects (Fig.~\ref{fig:DFT}(b)), where we use a lateral supercell to isolate the oxygen interstitials. For our other mechanism of surface exchange inequivalency, we choose as a representative substrate $(111)$-oriented $\mathrm{BaF_2}$ (Fig.~\ref{fig:DFT}(c)) due to its good lattice match with $(001)$ MBT (MBT must be strained to $\mathrm{BaF2}$ by only $2\%$), as well as the existence of prior experimental work using $(111)$ $\mathrm{BaF_2}$ as substrate for molecular beam epitaxial growth of $(\mathrm{MnBi_2Te_4})(\mathrm{Bi_2Te_3})$ thin films~\cite{Kagerer2020}.\\
\indent For our DFT calculations, we use the Vienna ab-initio software package (VASP)~\cite{Kresse1996}. We employ the generalized gradient approximation (GGA) with the Perdew-Burke-Ernzerhof (PBE) functional~\cite{Perdew1996}, and use the Projector-augmented wave (PAW) method\cite{Blochl1994} with standard VASP pseudopotentials. To approximately account for the localized nature of the unpaired $d$ electrons in $\mathrm{Mn}$, we use the DFT+U method~\cite{Anisimov1997} and adopt the rotationally invariant method by Dudarev et al.~\cite{Dudarev1998}. We set $U=4~\mathrm{eV}$ on the Mn atoms. Given that MBT is a layered material, we implement van der Waals corrections using the DFT-D3 method of Grimme et al.~\cite{Grimme2010}. We relax the bulk, primitive rhombohedral cell (doubled in the out-of-plane direction to accomodate the AFM order) until forces on all atoms are less than $0.01\mathrm{eV}/\mathrm{\AA}$. We use a Gamma-centered mesh of $13\time13\times3$ to sample the Brillouin zone and
a kinetic energy cutoff of $800~\mathrm{eV}$ for our plane-wave basis set. We neglect spin-orbit coupling (SOC), and enforce in-plane ferromagnetic order, and out-of-plane AFM order of the Mn ions using spin-polarized collinear calculations. We use a Gaussian smearing for the partial occupancies with a smearing width of $0.05~\mathrm{eV}$.\\  
\indent For the slab structures, we first convert the relaxed rhombohedral cell to the hexagonal setting and construct a $(001)$-oriented slab with six septuple layers and $20\mathrm{\AA}$ of vacuum. Although all DFT calculations are performed using a six-layer slab, we note that, provided that the slab is thick enough to avoid interactions between top and bottom surfaces, the surface exchange anisotropies $x$ should not depend on slab thickness. Thus, the results can be extrapolated to MBT crystals of both fewer and more layers.\\
\indent For simulations of ``full coverage" oxidation of one surface of MBT, we determine the lowest-energy position of the oxygen interstitial by comparing total energies of relaxed slab structures for several symmetry-inequivalent oxygen positions near one vacuum-terminated surface. We find that the energetically preferred $\mathrm{O}$ interstitial site sits between $\mathrm{Te}$ atoms just above the $\mathrm{Mn}$ layer closest to the surface, as shown in Fig.~\ref{fig:DFT}(a); this is consistent with ab-initio calculations in Ref.~\cite{mazza_surface-driven_2022}. In the relaxations for full-coverage oxidation, we use a slab with one $\mathrm{Mn}$ per layer in the unit cell, fix the lattice constant to the relaxed bulk values, and allow positions to relax until again the forces on all atoms are less than $0.01$ $\mathrm{eV}/\mathrm{\AA}$. The k-mesh is $11\times 11\times 1$. For the dilute limit of oxidation, where a $3\times 3\times 1$ supercell is used with one oxygen inserted, we initialize the structure with the same interstitial position found to be lowest energy in the full coverage case (Fig.~\ref{fig:DFT}(b)). Due to the very large number of atoms in the supercell ($379$, including the single oxygen defect) we relax the structure very roughly, until forces on all atoms are below $0.05$ $\mathrm{eV}/\mathrm{\AA}$ using a $3\times3\times1$ mesh, a $500~\mathrm{eV}$ kinetic energy cutoff.\\
\indent Finally, for the $(111)$ $\mathrm{BaF_2}$-$(001)$ $\mathrm{MnBi_2Te_4}$ heterostructure, we first relax the bulk, cubic unit cell of $\mathrm{BaF_2}$ with an $8\times8\times8$ k-mesh and $800~\mathrm{eV}$ kinetic energy cutoff until forces are less than $0.01\mathrm{eV}/\mathrm{\AA}$. We then convert the cubic cell to a hexagonal $(111)$-oriented setting. We next strain an isolated slab of MBT to the in-plane lattice parameters of $(111)$ $\mathrm{BaF_2}$, and relax positions and the out-of-plane lattice vector of MBT, while keeping the strained in-plane lattice vectors constant. Lastly, we create a heterostructure with three layers of $\mathrm{BaF_2}$ and six MBT layers with vacuum termination on each end (Fig.~\ref{fig:DFT}(c)), and relax atomic positions until all forces are less than $0.03\mathrm{eV}/\mathrm{\AA}$. We tested total energies of several interfacial lateral alignments, and found that the lateral alignment with the $\mathrm{Te}$ interfacial atoms of MBT directly above the $\mathrm{Ba}$ atoms, as shown in Fig.~\ref{fig:DFT}(c) is most energetically favorable.\\
\indent For all three situations, we compute the top-most and bottom-most interlayer AFM exchanges, labeling exchange at the modified surface (either due to oxygen interstitials or via interfacing to $\mathrm{BaF_2}$) as $(1+x)J^{\mathrm{AFM}}$, and exchange at the unmodified, vacuum-terminated surface as $J^{\mathrm{AFM}}$. From this we infer the surface exchange anisotropy $x$. The Heisenberg exchanges are calculated using the so-called ``four-state method"\cite{Xiang2011}, in which a given exchange between spins sites $i$ and $j$ is computed by taking total energy differences between four configurations with $(i,j)=(\uparrow,\uparrow);(\downarrow,\downarrow);(\uparrow,\downarrow);(\downarrow,\uparrow)$, and magnetic moments on all other sites in the unit cell held constant. For the pristine MBT structure, a given $\mathrm{Mn}$ atom in one layer has symmetry-degenerate AFM exchanges with three $\mathrm{Mn}$ atoms in the neighboring layer. In cases where the relaxation leads to inequivalent bond lengths, and thus slightly inequivalent Heisenberg couplings, we take in all cases the average of the three interlayer couplings. For the dilute limit of surface oxidation, we compute the average interlayer exchange for the surface $\mathrm{Mn}$ atom which is bonded to the oxygen; in the other two cases, all $\mathrm{Mn}$ at the modified surface have the same average interlayer exchange. We mention here that our calculated, absolute value of $J^{\mathrm{AFM}}$ for a pristine, vacuum-terminated surface is $0.1~\mathrm{meV}$, a smaller value, but on the same order of magnitude, as that found in Ref.~\cite{Otrokov2019B} and used in the MC simulations.\\
\indent We find that $x=-0.2$; $x=-0.07$; and $x=+0.2$ for full-coverage oxidation, diluate oxidation, and substrate growth on $(111)$ $\mathrm{BaF_2}$ respectively. Our results indicate that oxidation, at least for MBT, generically weakens the AFM interlayer coupling, whereas at least for the particular substrate of $\mathrm{BaF_2}$, contraction of the $\mathrm{Mn}$-$\mathrm{Mn}$ interlayer distance close to the interface leads to strengthening of the AFM interlayer coupling. These values are above the magnitude required to produce deterministic domain selection in thin-film MC simulations (see SI Sec.~J) and for the $x<0$ case of oxidation, are within a factor of two of the $x=-0.4$ value used for the bulk-like slab in the main text. Given the limited MC statistics, finite simulation size, and simplified spin Hamiltonian used in the MC simulations, we view this level of agreement as strong evidence that the phenomenological asymmetries required in our model are experimentally plausible for MBT.

\subsection{Dependence of domain selection robustness on single-ion anisotropy} The specific material of focus in our work, namely even-layer MBT, has a rather large single-ion magnetic anisotropy $K$ relative to its interlayer AFM coupling ($K/J^{\mathrm{AFM}}\approx0.5$), which is related to its quasi-two dimensional nature. This can be contrasted with more traditional AFMs where the single-ion anisotropy can be a few orders of magnitude smaller that AFM Heisenberg exchange. Therefore, to assess the range of our model's validity with respect to the relative values of single-ion anisotropy and interlayer AFM exchange, we perform MC simulations of a 48-layer MBT slab with the Heisenberg exchange parameters fixed to the values used in the main text, obtained from Refs.~\cite{Otrokov2019,Otrokov2019B}, while varying the value of single-ion anisotropy $K$ from $0.01$ to $1.5$ times the DFT-based value of $K=0.225~\mathrm{meV}$ used in the main text. For concreteness we refer to this reference value going forward as $K^*$. As a reminder, the fixed effective interlayer AFM coupling is $J^{\mathrm{AFM}}=0.47\mathrm{meV}$. As in the main text, we select a surface exchange anisotropy of $x=-0.4$ for the top surface and hold this value fixed.\ 
\begin{figure}
    \centering
   \includegraphics{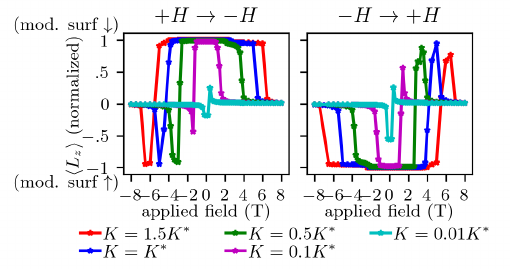}
   \caption{Behavior of field-based AFM domain selection as a function of single-ion anisotropy $K$. Panels (a) and (b) show MC simulation results for ramping-down and ramping-up protocols respectively, with a 48-layer MBT slab and a fixed surface exchange asymmetry of $x=-0.4$, as in the main text. Legend specifies the different $K$ values used as a fraction of $K^*=0.225\mathrm{meV}$ which was used in the main text based on the prior DFT calculations in Refs.~\citenum{Otrokov2019,Otrokov2019B}.}
   \label{fig:anisotropy}
   \vspace{0.5cm}
    \centering
   \includegraphics{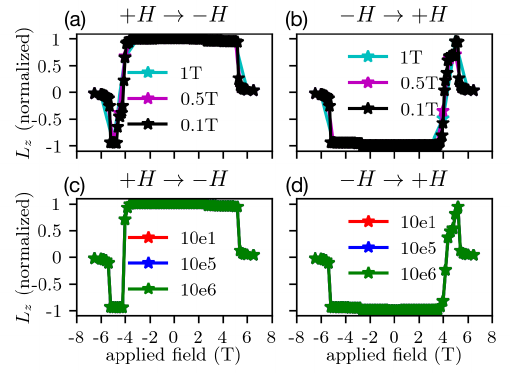}
   \caption{Behavior of field-based AFM domain selection for different effective ramping rates. All MC simulations are performed with the same 48-layer MBT slab and fixed surface exchange asymmetry Figs. (a) and (b) show ramping-down and ramping-up protocols respectively for simulations with varying step sizes between successive field values in the MC simulations, indicated in the legend. Figs. (c) and (d) show ramping-down and ramping-up simulations respectively for varying numbers of MC equilibration steps performed at each field value before evaluation of magnetic properties.}
   \label{fig:ramprate}
\end{figure}
\indent The results are shown in Fig.~\ref{fig:anisotropy}., where we have plotted the ensemble-averaged Néel vector as a function of applied field for (a) ramping down and (b) ramping up protocols. We see that at $H_z=0$ we get $100\%$ selection of the expected domain for a given ramping direction, even when the relative uniaxial anisotropy reduced to $K=0.1K^*$ times the value of the main text (corresponding to $0.0225~\mathrm{meV}$ and $K/J^{\mathrm{AFM}}\approx0.05$). However, the plots also reveal a clear difference as a function of $K$ in that the critical field at which the intermediate ``bilayer" state with pinned surface layers transitions to a monodomain AFM state decreases with decreasing $K$. This is due to the fact that the energetic penalty of canting away from the direction of the field decreases as the magnitude of easy-axis anisotropy decreases, thus energetically stabilizing the spin-flop phase relative to the collinear multidomain bilayer phase for a wider range of magnetic fields. By the time we get to the weakest tested anisotropy value $K=0.01K^*$, inspection of the MC data reveals that the slab remains in the spin-flop phase until arbitrarily close to zero field, and the following monodomain AFM phase at $H=0$ is random., as seen in the negligible values of the ensemble-averaged $\left<L_z\right>$ in Fig.~\ref{fig:anisotropy}.\\
\indent Fig.~\ref{fig:anisotropy} highlights an essential criterion for our model of field-based deterministic domain selection to remain valid. For a fixed temperature and field range, the uniaxial anisotropy must be strong enough to stabilize the BL phase in an intermediate field window between the high-field spin-flop phase and low-field mondomain AFM state. If the anisotropy becomes too weak relative to the interlayer AFM exchange, one reaches a regime where the bilayer phase is never the lowest energy state, and the direct evolution from the spin-flop to monodomain phase is in this case random.

With all other ab-initio derived magnetic parameters for MBT fixed to those used in the main text, at a constant temperature of $15\mathrm{K}$ (also the temperature at with the MOKE measurments in Ref.~\cite{Sunko2026} were performed), domain selection remains deterministic down to $K/J^{\mathrm{AFM}}\approx0.05$. Thus, our proposed mechanism of deterministic domain selection based on inquivalent surface exchanges is robust for a fairly wide range of anisotropies, assuming all other parameters fixed. At the same time, the weakening of deterministic domain selection as the uniaxial anisotropy is further lowered, together with the general qualitative arguments above, also indicate that this works' findings are most robust for AFMs with moderately large anisotropy relative to their AFM exchange. MBT is just one example of such an AFM; high anisotropy relative to isotropic exchange is a general feature of low-dimensional, van der Waals AFMs, since in reduced dimensions large anisotropy is often necessary to stabilize magnetic order. On the other hand, for a truly three-dimensional bulk AFMs where single-ion anisotropy can be on the order of $\mu\mathrm{eV
}$ in contrast to $
\mathrm{meV}$ dominant AFM coupling, deterministic domain selection based on our proposed mechanism may be more difficult to achieve.\\
\indent We emphasize in closing however that the robustness of deterministic domain selection via our model does not only depend on the relative values of $K$ and $\mathrm{meV}$, but instead on the overall balance between Zeeman energy, thermal energy, anisotropy,exchange, as well as the magnitude of top and bottom surface exchange inequivalence. Thus, for a fixed $K/J^{\mathrm{AFM}}$, one can strengthen or weaken the deterministic nature of domain selection by tuning temperature, or by scaling the absolute values of other paramters. For example, we find that by keeping $x=-0.4$ but uniformly scaling both intralayer and interlayer exchanges in our simulation of MBT, as well as the range of applied fields, by a factor of 10, and taking $K$ such that $K/J^{\mathrm{AFM}}=0.005$, at $15\mathrm{meV}$ the domain selection is fully deterministic, in contrast to the same ratio in Fig.~\ref{fig:anisotropy}, corresponding to $K=0.01K^*$. This is because increasing the overall magnetic exchange strength increases the AFM ordering temperature, meaning thermal fluctuations are relatively suppressed at a fixed temperature. Similarly, one can keep the exchange and anisotropy magnitudes fixed while lowering the absolute temperature. Thus, while it is clear that our proposed mechanism is valid for the widest range of experimental conditions for AFMs with relatively large anisotropy, deterministic domain selection for small-anisotropy sytems may still be possible if other parameters, in particular temperature, are carefully tuned.
\subsection{Dependence of domain selection robustness on field ramping rate}
Deterministic AFM domain selection via the mechanism proposed in this work has two steps. The first is the thermodynamic selection of lowest-energy bilayer states at intermediate fields due the combination of surface pinning and surface exchange inequivalence. The second is the subsequent degeneracy breaking of pathway probabilities (as analytically validated in our committor analysis) to the two energetically degenerate AFM monodomains, starting from the surface-localized bilayer state. Given especially that the second step is pathway-dependent, it is relevant to consider how robust our proposed mechanism remains under different ramping rates.
\indent While experimental field ramping rates cannot be rigorously captured in Monte Carlo simulations, which do not have a true timescale, we can still approximate the experimental control knob of ramp rates in a few ways. Specifically, we tested the robustness of the selection by varying the field increment between successive Monte Carlo simulations, or changing the number of Monte Carlo equilibration steps (before evaluation of system properties begins) at each field value. These tests roughly model experimental situations in which the increments between successive applied fields vary, and situations in which each field value is held constant for different time periods, respectively. In Figs. 3 and 4(a) of the main text, as well as in the other SM MC simulations, we have varied the field in steps of $0.2\mathrm{T}$ between $-4\mathrm{T}$-$+4\mathrm{T}$, and $0.5\mathrm{T}$ at greater field magnitudes, and we have performed $500000$ equilibration steps before ``measurement" at each field value.

\indent In Figs.~\ref{fig:ramprate}(a)-(b) we show ramp-down and ramp-up MC simulations respectively where we keep the number of MC equilibration steps at $5\mathrm{e}5$ and have varied the step size between successive field values from starting points of $\pm6.5\mathrm{T}$, from a step size of $0.01~\mathrm{T}$ all the way up to a step size of $1~\mathrm{T}$. In Figs.~\ref{fig:ramprate}(c)-(d) we show ramp-down and ramp-up simulations respectively where the step size is fixed at $0.2~\mathrm{T}$ but the number of equilibration steps at each field value ranges from $10$ to one million. In all four panels, the curves for all situations, including $1\mathrm{T}$ step sizes and $10$ equilibration steps, which are numerical proxies for rapid ramp rates, lie almost exactly on top of each other. Although we emphasize again that MC steps cannot be converted into a rigorous physical ramp rate, these results strongly suggest that the proposed selection mechanism does not require an experimentally fine-tuned ramp protocol, and should be robust for a wide range of ramping rates.

\subsection{Scaling of deterministic domain selection with number of layers}
In our model, deterministic evolution to a specific AFM domain is dictated by whether, in the intermediate-field regime, the system enters a bilayer phase with the ferromagnetically coupled layer pair localized at one of the two surfaces. In this case, evolving into the one of the two monodomains as the field is further lowered will require only a single layer flip, whereas evolving into the opposite domain requires reversal of the remaining N-1 layers, or equivalently, propagation of the domain wall across the entire slab, and is thus vanishingly likely. Therefore, in the scaling arguments below, we focus exclusively on the probability of transitioning from the spin-flop phase to the \textit{more likely} of the two surface-localized bilayer states, compared to transitioning to any of the other $\frac{N}{2}-1$ field-stabilized bilayer positions.\\
We assume a large number $N_{\parallel}$ of in-plane, ferromagnetically coupled spins in each layer. We also assume that the intraplanar FM coupling is sufficiently larger than $J^{\mathrm{AFM}}$, such that all spins in each layer are aligned parallel below the AFM transition temperature; this is the case for MBT and other layered van der Waals AFMs. Then, the total energy difference between the state with the bilayer localized at the exchange-modified surface (``mod") and a state with any of the other energetically degenerate $\frac{N}{2}-1$ bilayer positions (``other") is
\begin{equation}
   \Delta E=E(\mathrm{BL,mod})-E(\mathrm{BL,other})=2J^{\mathrm{AFM}}xN_{\parallel}
   \label{eq:BL_E_diff}
\end{equation}
Recall that with the convention of the Hamiltonian in the main text, $J^{\mathrm{AFM}}$ is positive. Thus, $\Delta E$ is negative (positive) when $x<0$ ($x>0$).\\
\indent We first consider an $N$-layer AFM with $x<0$, In this case, the state with bilayer position at the exchange-modified surface is lower in energy by $2J^{\mathrm{AFM}}|x|N_{\parallel}$ compared to the $\frac{N}{2}-1$ other bilayer positions which are energetically degenerate. We exclude bilayer positions where the bilayer spins are antiparallel to the field, since these are never energetically stabilized with respect to the spin-flop or monodomain phases. The relative probability to transition from the spin-flop phase to \textit{any} of the higher-energy bilayer states, compared to the transition probability to the single lower-energy bilayer state for a given temperature $T$ is given by
\begin{equation}
\begin{aligned}
\epsilon
&= \frac{P(\mathrm{SF}\rightarrow\mathrm{BL,other})}
        {P(\mathrm{SF}\rightarrow\mathrm{BL,mod})} \\
&= \left(\frac{N}{2}-1\right)
   \exp\left(
      -\frac{2|x|N_{\parallel}J^{\mathrm{AFM}}}{k_{\mathrm B}T}
   \right).
\end{aligned}
\label{eq:BL_prob}
\end{equation}
\indent To get a more useful expression, we assume $N$ is fairly large such that $\frac{N}{2}-1\approx\frac{N}{2}$. We also choose a threshold probability ratio $
\epsilon^*$ above which we say that the probability to reach one of the higher-energy BL states is high enough for the domain selection process to be non-deterministic. Then, by rearranging we can get a critical value of surface exchange anisotropy $|x^*|$ as a function of layer number $N$:
\begin{equation}
    |x^*| \sim 
    \frac{k_B T}{2N_{\parallel}J^{\mathrm{AFM}}}
    \ln\left(\frac{N}{2\epsilon}\right).
    \label{eq:xcrit}
\end{equation}
\indent Thus, the minimal value of surface exchange anisotropy to achieve deterministic domain selection for $x<0$ increases logarithmically, rather than linearly, with the thickness, or number of layers $N$ of the AFM. This already indicates that the required surface anisotropy does not become arbitrarily large as $N$ approaches a bulk-like limit.\\
\begin{figure*}
    \centering
   \includegraphics{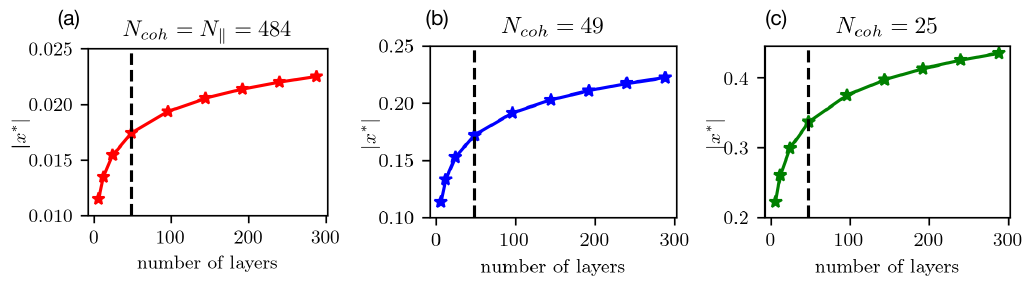}
    \caption{Critical absolute value of surface exchange anisotropy $|x^*|$ (Eq.~\ref{eq:xcrit}) for deterministic domain selection when $x<0$ as a function of AFM layer number $N$, assuming the transition from the spin-flop to a particular bilayer state occurs once $N_{coh}$ ferromagnetically-coupled spins in each layer start to align in the bilayer configuration. (a) assumes $N_{coh}$ is equal to the in-plane number of parallel spins $N_{\parallel}=484$ in our Monte Carlo simulations (i.e. precisely Eq~\ref{eq:xcrit}, while (b) and (c) replace $N_{\parallel}$ in Eq.\ref{eq:xcrit} with smaller numbers of $N_{coh}=49$ and $N_{coh}=25$ respectively. We have set the cutoff for deterministic selection at $\epsilon=0.056$ in these plots, and the dashed line at $N=48$ indicates the thickness used for our MC simulations in the main text.}
    \label{fig:scaling}
\end{figure*}
\begin{figure}
    \vspace{0.5cm}
    \centering
   \includegraphics{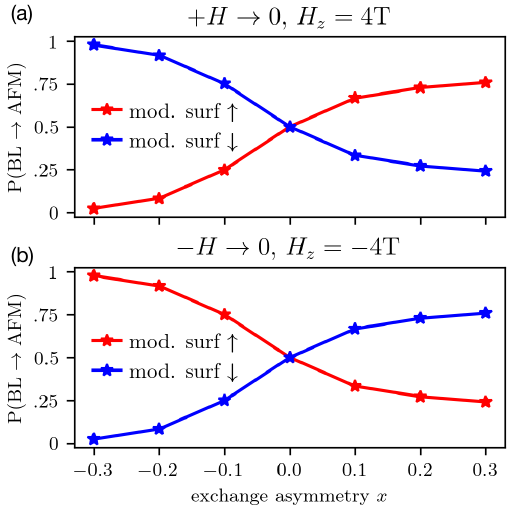}
    \caption{Total transition probabilities of a six-layer slab for varying surface exchange asymmetry $x$ from any BL state to either domain ``mod. surf $\uparrow$" or ``mod. surf $\downarrow$". (a) evaluated at $+4~\mathrm{T}$ reflects a ramping-down field and (b) at $-4~\mathrm{T}$ represents ramping-up.} 
    \label{fig:statmech}
\end{figure}
\indent Choosing $\epsilon=0.0526$ (corresponding to $P(\mathrm{SF}\rightarrow\mathrm{BL,other})=0.05$), in Fig. 
\ref{fig:scaling}(a) we plot $|x^*|$ as a function of $N$ with all other parameters in Eq. \ref{eq:xcrit} held constant to the values of Figs. 3-4 in the main text: $J^{\mathrm{AFM}}=0.47\mathrm{meV}$, $T=15\mathrm{K}$, and $N_{\parallel}=484$. The dashed vertical line at $N=48$ shows the thickness used in our Monte Carlo simulations. Interestingly, this figure suggests that the critical value of surface exchange anisotropy required to achieve deterministic selection is tiny, $x\approx-0.02$. This contradicts our finding that we only get completely deterministic selection for this thickness around $x=-0.4$. Given this, we adjust the implicit assumption in Eqs. \ref{eq:BL_prob} and \ref{eq:xcrit}, namely that all $N_{\parallel}$ in-plane spins in each layer act as a rigidly rotating macrospin when transitioning from the spin-flop to BL phase. Instead, we hypothesize that smaller clusters of $N_{coh}<N_{\parallel}$ FM-coupled spins in each layer coherently rotate to a specific BL state.\\
\indent Figs. \ref{fig:scaling}(b)-(c) show $|x^*|$ based on Eq.\ref{eq:xcrit} with $N_{coh}$ replacing $N_{\parallel}$ for $N_{coh}=49$ and $N_{coh}=25$, and we see that for $N_{coh}=25$ in Fig.~\ref{fig:scaling}(c) $|x^*|$ is between $0.3$ and $0.4$, in very good quantitative agreement with our MC calculations. Of course, given the approximations and inaccuracies inherent in the MC calculations, the actual $N_{coh}$, and corresponding $|x^*|$, could be slightly smaller or larger than than what we find.\\
\indent The main takeaway message, regardless of the exact prefactor, is that when $x<0$, the critical anisotropy for deterministic domain selection stays at experimentally plausible values, on the order of magnitude of those found in our DFT calculations, up to hundreds of layers, well within the bulk-like regime for MBT, due to the logarithmic scaling.\\
\indent When neither surface exchange is weakened with respect to the bulk
$J^{\mathrm{AFM}}$, the story is completely different. We again take the case
of Fig.~2 in the main text, where one surface retains the bulk
$J^{\mathrm{AFM}}$ coupling and the other has a stronger AFM coupling,
$(1+x)J^{\mathrm{AFM}}$ with $x>0$. The BL localized at the modified
surface is then higher in energy than the remaining $\frac{N}{2}-1$
BL positions, and is therefore vanishingly unlikely. The only probable state with a surface-localized BL
leading to deterministic AFM domain selection is instead the one at
the opposite surface.\\
\indent Crucially, this opposite-surface BL is not energetically selected by increasing $x$: it remains degenerate with the $\frac{N}{2}-2$ central
BL states. The central BL states give canceling
committor contributions to the two AFM domains, so the net domain
bias for $x>0$ comes only from the
surface-localized BL state which is energetically degenerate with these central states. Assuming equal accessibility of all degenerate
BL states from the spin-flop phase, this weight is diluted by the
growing central-BL manifold and scales as $\sim 1/N$. Moreover, since for $x>0$, $x$ governs the energy of the higher-energy BL states, which in any case is vanishingly unlikely, there is
no critical $|x^*|$ for deterministic selection analogous to the $x<0$ case.\\
\indent In fact, this equal-accessibility estimate underestimates the degree of
selection seen in MC simulations of thin-film six-layer MBT which we describe further in SI Sec.~J. In Fig.~\ref{fig:MC_sixlayers}, the
tested $x>0$ parameters give full selection of the expected domain,
whereas the six-layer committor analysis (Fig.~\ref{fig:statmech} in SI Sec.F) gives a strong but non-unit
bias. The likely reason is kinetic: during the spin-flop-to-BL
transition, we see from inspection of the MC simulations that for all slab thicknesses, the surface moments are already nearly parallel to the
field at field values for which the central layers remain canted in the spin-flop phase. This likely makes surface-BL
nucleation more accessible than central-BL nucleation, despite energetic degeneracy of the final collinear BL states states. This kinetic
preference appears to dominate when few central BL states compete. It does
not, however, change the $\frac{1}{N}$ scaling conclusion for $x>0$: in the
48-layer MBT simulations, the growing number of energetically degenerate central BL states
is seen to outweigh the kinetic surface-BL preference, and we do not obtain fully
deterministic selection for any tested $x>0$, though we do for $x<0$.\\
\subsection{Committor analysis}
In the main text, we argued qualitatively, and validated with Monte Carlo simulations, that the evolution from the intermediate-field bilayer phase to an AFM monodomain phase as $H_z\rightarrow 0$ can be deterministic when surface interlayer exchanges are inequivalent. We further quantify these arguments here via ``committor functions", which give the probability that, starting from an initial state A, we end up in possible final state B before hitting possible final state C~\cite{Lindner2019}. We assume sufficiently large uniaxial anisotropy such that all states at fields below $|{H_z^{SF}}|$ are collinear, giving $2^{N}$ configurations, including the $\frac{N}{2}$ possible BL states and the two AFM domains. We model evolution at fixed $H_z$ and temperature $T$ as a Markov process where a transition from state $i$ to $j$ has probability
\begin{equation}
P(i\to j) =
\begin{cases}
\frac{1}{N}\min\bigl(1, e^{-\frac{E_j - E_i}{K_bT}}\bigr), & d_{\mathrm{H}}(i,j)=1,\\[6pt]
0, & \text{otherwise},
\end{cases}
\label{eq:Transprobs}
\end{equation}
where $d_{\mathrm{H}}(i,j)=1$ indicates that $i$ and $j$ differ by one spin flip, and the energies are evaluated via Eq.~1 in the main text.\\
\indent The committor function to transition from BL state $a$ to collinear AFM domain $A$, which implicitly takes into account probabilities of all possible transition paths of arbitrary length, is given by
\begin{equation}
    q_{\mathrm{AFM},A}(\mathrm{BL},a)=[(\mathcal{I}-\mathcal{Q})^{-1}\mathcal{R}]_{aA}.
    \label{eq:committor}
\end{equation}
$\mathcal{I}$ is an identity matrix and $\mathcal{Q}$ (size $(2^N-2)\times(2^N-2)$) contains the one-step transition probabilities between all intermediate collinear configurations, \textit{excluding} the two AFM domain states. Diagonal elements enforce total transition probabilities from each state to sum to unity. $\mathcal{R}$ (size $(2^N-2)\times2$) contains the one-step transition probabilities from the $(2^N-2)$ ``intermediate" states to one of the two AFM domain states. The row index of Eq. \ref{eq:committor} selects the BL state and the column index the AFM domain. To find the total probability of transitioning from any BL state, we compute a Boltzmann-weighted sum:
\begin{equation}
    \mathrm{P}(\mathrm{BL}\rightarrow\mathrm{AFM},A)=\sum_{\mathrm{BL},i}\frac{e^{-E_{\mathrm{BL},i}/K_BT}q_{\mathrm{AFM},A}(\mathrm{BL},i)}{\sum_{\mathrm{BL},j}e^{-E_{\mathrm{BL},j}/K_BT}},
\label{eq:BL_weight}
\end{equation}
which runs over the $N/2$ BL states.\\
\indent For our toy model calculations we choose $J^{\mathrm{AFM}}=3~\mathrm{meV}$, $|S|=4.5~\mu_B$, and $K=1~\mathrm{meV}$: because our toy model involves only one spin per layer, the values of exchange and anisotropy have to be significantly increased with respect to those of layered van der Waals AFMs with many spins per layer in order to get comparable results in terms of the magnitudes of surface exchange anisotropy $x$. We evaluate the committors at $T=5~\mathrm{K}$ and fields of $+4~\mathrm{T}$ ($-4~\mathrm{T}$) to represent ramping-down (ramping-up) protocols; the fields were chosen below the values where the collinear AFM states become lowest in energy, (see SI, Sec.~K). In Fig.~\ref{fig:statmech} we plot both $\mathrm{P}(\mathrm{BL}\rightarrow$ mod.~surf $\uparrow)$ and $\mathrm{P}(\mathrm{BL}\rightarrow$ mod.~surf $\downarrow)$ versus $x$ for (a) ramping-down and (b) ramping-up protocols. Results are consistent with our qualitative arguments; the AFM domain corresponding to the surface spin with \emph{weaker} AFM exchange pointing \emph{opposite} the initial field sign is consistently favored.\\ 
\indent Further, for fixed $|x|$,  bias is stronger for $x<0$ and weaker for $x>0$. This is a consequence of the fact discussed in the main text, and in SI Sec.~E, that deterministic selection in the $x<0$ limit is more robust than for $x>0$.
\subsection{Dependence of domain selection robustness on magnetic exchange, thickness and temperature}
\begin{figure*}
    \centering
   \includegraphics{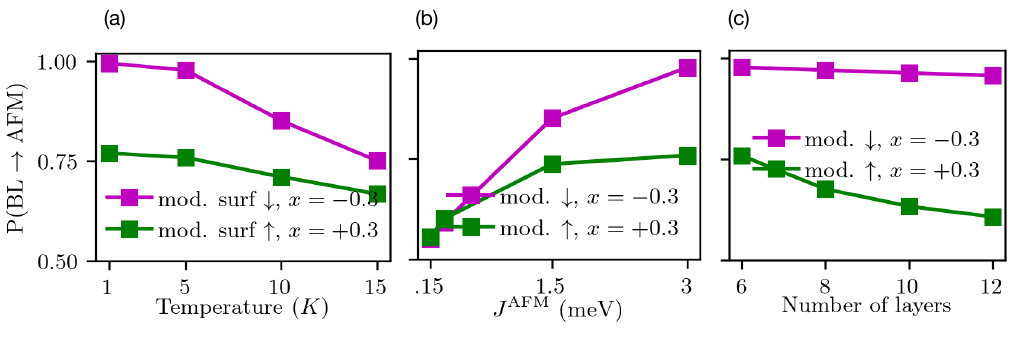}
   \caption{Committor functions showing summed probability to transition from any of the bilayer states to one of the two AFM domains as function of (a) temperature; (b) $J^{\mathrm{AFM}}$ and (c) number AFM-coupled layers in the slab. We show results for $x<0$ (purple) and $x>0$ (green) with $|x|=0.3$, and we only plot the committor the AFM domain which is favored for each sign of $x$ for the ramping down protocol assumed in the calculations. (a) and (b) are calculated with a six-layer slab. The temperature is $5\mathrm{K}$ when it is not being varied, and $J^{\mathrm{AFM}}=3\mathrm{meV}$ when not being varied. Uniaxial anisotropy $K$ and magnetic moment value $S_i=4.5\mu_B$ are kept constant in all cases.}
   \label{fig:prob_param_dependence}
    \vspace{0.5cm}
   \includegraphics{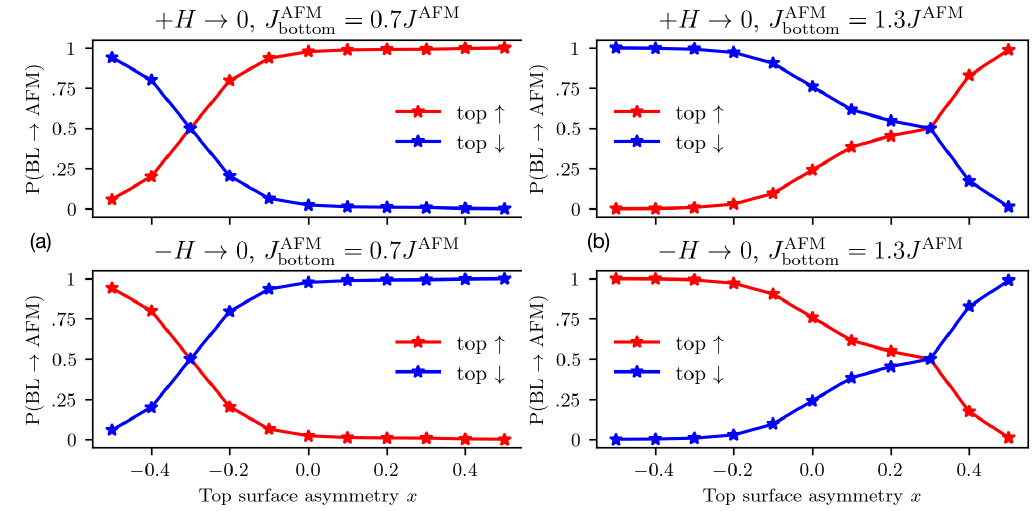}
   \caption{Committor functions showing summed probability to transition from any of the bilayer states to one of the two AFM domains as function of top layer surface asymmetry $x$. All interior interlayer couplings are set to $J^{\mathrm{AFM}}=3~\mathrm{meV}$, just like Fig.~\ref{fig:statmech} in SI Sec.~F. In contrast to SI Sec.~F, here the bottom surface exchange is also different from the bulk interlayer couplings, such that in (a) it is fixed at $x_{\mathrm{bottom}}=-0.3$ ($J^{\mathrm{AFM}}_{\mathrm{bottom}}=0.7J^{\mathrm{AFM}}$) and in (b) it is fixed at $x_{\mathrm{bottom}}=+0.3$ ($J^{\mathrm{AFM}}_{\mathrm{bottom}}=1.3J^{\mathrm{AFM}}$).}
   \label{fig:comm_bothsurf}
\end{figure*}

It is important to understand how the robustness of deterministic domain selection for fixed surface exchange asymmetry and ramping direction changes when other parameters are varied. In Fig. \ref{fig:prob_param_dependence}, we evaluate the committor functions for transitioning from the intermediate bilayer (BL) phase to either of the two collinear AFM domains (Eq.~\ref{eq:BL_weight}) where we vary (a) temperature; (b) $J^{\mathrm{AFM}}$; and (c) slab thickness. We show results for both $x<0$ (purple) and $x>0$ (green) exchange anisotropy of the top surface with a consistent magnitude $|x|=0.3$. The positive magnetic field at which the energies going into the commitor are calculated is $H_z=+4~\mathrm{T}$ for $J^{\mathrm{AFM}}=3\mathrm{meV}$ (the same field magnitude used to calculate the committors in the previous section) and it is scaled proportionally when $J^{\mathrm{AFM}}$ is varied in Fig. \ref{fig:prob_param_dependence}(b). We show only the domains which are favored for this ramping direction. They correspond to the mod. surf $\downarrow$ domain for $x=-0.3$ and the mod. surf $\uparrow$ domain for $x=+0.3$.\\
\indent For all three plots, the trend is the same for both signs of anisotropy, but the details vary. We first consider the committors as a function of varying temperature in Fig. \ref{fig:prob_param_dependence}(a). For both $x<0$ and $x>0$ the probability to transition to the favored domain decreases with increasing temperature since transitions to the lower-energy states are suppressed by the Boltzmann factor $\exp \left(1/K_BT\right)$, both in the intermediate, single spin flip transitions (Eq. \ref{eq:Transprobs}) and in the overall weighted committor (Eq. \ref{eq:BL_weight}). However, the effect is stronger in the $x<0$ case. We can understand this by noting that for $x<0$, Boltzmann weight in Eq. \ref{eq:BL_weight} corresponding to the BL state with the bilayer at the surface with modified exchange dominates over all other BL states. For $x>0$ on the other hand, $\frac{N}{2}-1$ possible BL positions with equal liklihood contribute significantly to the overall transition probability, so the temperature dependence in Eq. \ref{eq:BL_weight} less strongly re-weights the energetically favorable and unfavorable BL states than in the $x<0$ case.
\indent In Fig. \ref{fig:prob_param_dependence}(b), we plot Eq. \ref{eq:BL_weight} as a function of interlayer AFM coupling $J^{\mathrm{AFM}}$, given in $\mathrm{meV}$. The transition probability for the favored domain increases with $J^{\mathrm{AFM}}$. This is because the energy lowering in going from ferromagnetically coupled bilayers to AFM-coupled layers increases with $J^{\mathrm{AFM}}$, and spin-flips leading to the AFM state (Eq. \ref{eq:Transprobs}) thus become more likely.\\
\indent In Fig. \ref{fig:prob_param_dependence}(c) we plot Eq. \ref{eq:BL_weight} as a function of number of out-of-plane layers (the matrix sizes are too large for efficient evaluation of the committor function beyond 12 layers). For both $x<0$ and $x>0$ the robustness of deterministic domain selection decreases with increasing number of layers. This is because the number of BL states with the BL, or domain wall, in the interior of the slab increases with increasing layer number. These states have similar probability to transition to either AFM domain, so they suppress the overall robustness of deterministic selection based on the field-ramping direction. For surface exchange asymmetry with $x<0$ however, the decrease in domain selection robustness is much weaker as a function of layer number, since the central bilayer states are higher energy and thus still relatively unlikely compared to the state where the BL is at the surface with the weaker exchange (See SI Sec.~E for more quantitative discussion). For $x>0$ on the other hand, all the central BL states are low-energy compared to the state with the BL at the more strongly AFM surface. Thus, an increase in these low-energy states with increasing layer number strongly decreases probability for the favored AFM domain to be selected.\\
\indent Finally, we mention that the robustness of deterministic domain selection probability also depends on the uniaxial single-ion anisotropy $K$. However, because we only consider collinear states in the Ising-like toy model used in our committor functions (in order to restrict the state space to a tractable size), evaluation of Eqs. \ref{eq:Transprobs}-\ref{eq:BL_weight} yields no information about the barriers for spin rotation set by $K$. Instead, to assess the dependence of domain selection robustness on uniaxial anisotropy, in SI Sec.~C we separately perform Monte Carlo simulations of MBT with varying values of $K$ relative to other fixed parameters.
\subsection{Committor functions for $J^{\mathrm{AFM}}_{\mathrm{bottom}}\neq J^{\mathrm{AFM}}_{\mathrm{bulk}}$}
In the above sections, we use a simplified model of surface exchange asymmetry where the bottom-most interlayer exchange between layers $1$ and $2$ keeps the bulk value $J^{\mathrm{AFM}}$, and only the top exchange between layers $N$ and $N-1$ varies. In reality, however, atoms at any vacuum-terminated surface or interface will typically relax, or even reconstruct, with respect to their bulk positions as a result of the altered chemical environment~\cite{Feibelman1996,Finnis1974}. Thus, in general \textit{both} top and bottom surface exchanges will be modified from the bulk $J^{\mathrm{AFM}}$. In Fig.~\ref{fig:comm_bothsurf} we calculate committor functions (Eq.~\ref{eq:BL_weight}) to transition from the BL phase to either the top surf $\uparrow$ (red, called mod. surf $\uparrow$ in main text) or top surf $\downarrow$ (blue, called mod. surf $\downarrow$ in the main text) AFM domain at a positive (ramp-down) or negative (ramp-up) $H_z$ value where the transition is likely to occur based on the relative energetics of BL states and AFM states. Ramp-down cases are shown in the top panels, and ramp-up cases are in the bottom panels. Analogously to Fig.~\ref{fig:statmech} in SI sec.~F, we plot the committors for both AFM domains as a function of varying top surface exchange asymmetry $x$, but now we fix the bottom interlayer exchange to either (a)$J^{\mathrm{AFM}}_{\mathrm{bottom}}=0.7J^{\mathrm{AFM}}$ (corresponding to $x_{\mathrm{bottom}}=-0.3$) or (b) $J^{\mathrm{AFM}}_{\mathrm{bottom}}=1.3J^{\mathrm{AFM}}$ ($x_{\mathrm{bottom}}=+0.3$). We use the same parameters as in the committor calculations of SI Sec. F, namely $J^{\mathrm{AFM}}=3~\mathrm{meV}$, $K=1~\mathrm{meV}$, a temperature of $5\mathrm{K}$, and applied fields of $+4~\mathrm{T}$ ($-4~\mathrm{T}$) for ramping-down (ramping-up) calculations.\\
\indent It is apparent in both cases that the preferred domain for a given ramping protocol follows the same \textit{relative} logic as for the situation of a single surface exchange differing from the bulk, shown in Fig.~\ref{fig:statmech}. Specifically, the domain with top surface spins pointing opposite (parallel) to the initial sign of $H_z$ in the ramping protocol is favored when the top surface exchange is weaker: $x_{\mathrm{top}}<x_{\mathrm{bottom}}$ (stronger: $x_{\mathrm{top}}>x_{\mathrm{bottom}}$) than the bottom surface exchange. Note that now, the crossover points where there is equal probability to transition to either domain occur not when the top surface $x$ is zero, but instead when $x_{\mathrm{top}}=x_{\mathrm{bottom}}$.\\ \indent Thus, our arguments in this work are robust for more realistic models of surface exchange asymmetry where $J^{\mathrm{AFM}}_{\mathrm{top}}\neq J^{\mathrm{AFM}}_{\mathrm{bottom}}$ and both surface exchanges differ from the bulk-like $J^{\mathrm{AFM}}$ of the central layers. The only caveat is that both $x_{\mathrm{bottom}}$ and $x_\mathrm{top}$ should lie within $-1$ and $+1$ such that the effective exchanges for both surfaces are less than the effective exchanges $2J^{\mathrm{AFM}}$ of central layers, causing the bilayer phase with pinned surface spins to be stabilized at intermediate fields.\\
\indent As a final comment, the asymmetry in deterministic domain selection for weaker, $x<0$, compared to stronger, $x>0$ surface exchange relative to the bulk interlayer $J^{\mathrm{AFM}}$ is also apparent in this more realistic surface model. In Fig. \ref{fig:comm_bothsurf}(a) where $x_{\mathrm{bottom}}=-0.3$, we already get the favored AFM domain for both ramping protocols with nearly unit probability when $x_{\mathrm{top}}=0$, implying a relative asymmetry between top and bottom surfaces of $0.3$. In Fig. \ref{fig:comm_bothsurf}(b) in contrast, where $x_{\mathrm{bottom}}=+0.3$, we do not get definitive domain selection until $x_{\mathrm{top}}\approx =-0.2$, implying that a relative surface asymmetry of $0.5$ is required. Interestingly, from Fig. \ref{fig:comm_bothsurf}(b), we see that once $x_{\mathrm{top}}$ is more positive than $x_{\mathrm{bottom}}$, the domain selection asymmetry grows very rapidly with increasing $x_{\mathrm{top}}$. From inspection of the the committor functions from individual bilayer states to specific domains (as opposed to the weighted sum over all possible bilayer states calculated in the figure), this appears to stem from a strongly suppressed probability for states with central bilayer positions to transition to the disfavored domain. This is because this transition requires an intermediate step where the bilayer moves to the top surface which has stronger AFM coupling than the bottom surface. We caution that we have not yet observed this phenomenon in our Monte Carlo simulations, so the trend should be taken with a grain of salt until follow-up MC simulations are performed with varying top and bottom surface exchanges.\\

\subsection{Monte Carlo simulations for 48-layer $\mathrm{MnBi_2Te_4}$ with other $x$ values }
\begin{figure}
    \centering
   \includegraphics{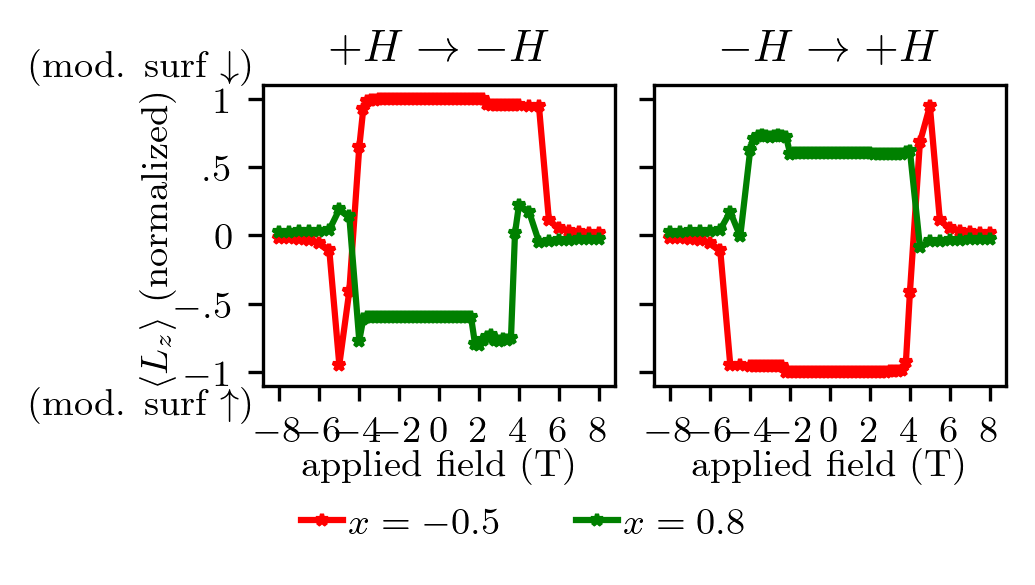}
   \caption{$\hat{z}$-component of normalized N\'{e}el vector $L$ as a function of applied field based on Monte Carlo (MC) simulations of 48-layer MBT for surface exchange asymmetries $x=-0.5$ and $x=0.8$.}
   \label{fig:MC_otherx}
    \vspace{0.5cm}
    \includegraphics{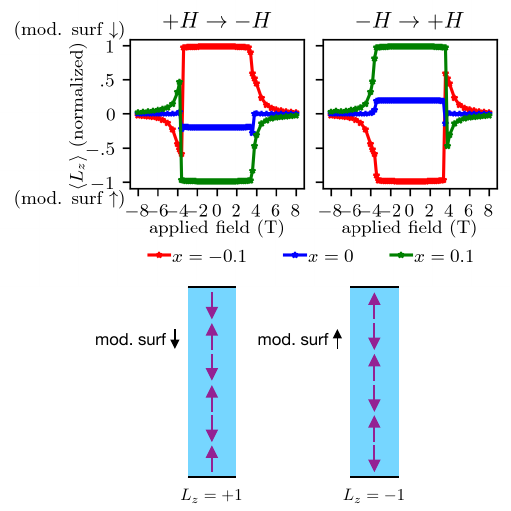}
   \caption{$\hat{z}$-component of ensemble-averaged normalized N\'{e}el vector $\langle L_z\rangle$ as a function of applied field based on Monte Carlo (MC) simulations for varying surface exchange asymmetry $x$ for a six-layer slab of MBT. Left-hand side is a ramping down simulation, and right-hand side is a ramping up simulation. Bottom cartoons show the correspondence between the sign of $\langle L_z\rangle$ and the ``mod. surf $\uparrow$" and ``mod. surf $\downarrow$" domains using our convention.}
   \label{fig:MC_sixlayers}
\end{figure}
In the main text, our goal was to demonstrate that finite asymmetry in magnetic exchange between top and bottom surfaces can lead to deterministic domain selection via magnetic fields as observed experimentally in Ref.~\cite{Sunko2026}. Thus, in Figs. 3 and 4 of the main text we only show the ensemble-averaged AFM order parameter $\langle L_z\rangle$ as a function of applied field for the smallest magnitude of surface exchange asymmetry ($x=-0.4$) which we found to lead to 100 percent selection of the same domain for a fixed ramping direction in all MC simulations. Here in Fig. \ref{fig:MC_otherx} we plot $\langle L_z\rangle$ versus $H_z$ for ramping-down (left) and ramping-up (right) for a larger negative value of asymmetry, $x=-0.5$, and for a positive surface exchange asymmetry $x=0.8$.\\
\indent Qualitatively, $\langle L_z\rangle$ vs. $H_z$ for $x=-0.5$ is very similar to the result for $x=-0.4$ in Fig. 4(a) of the main text. Transitions between spin-flop and multidomain BL states occur at $H^{SF}=\pm5.25~\mathrm{T}$. For ramping-down, the transition from the energetically favorable BL $\uparrow$ state (with the BL at the surface with weaker exchange) the mod. surf $\downarrow$ domain occurs at $2.2~\mathrm{T}$, which is slightly lower than the $2.6~\mathrm{T}$ value for $x=-0.4$. This is because the larger value of $x$ in Fig. \ref{fig:MC_otherx} means that the ferromagnetically coupled BL is energetically favored compared to AFM interlayer coupling until lower applied field values, so the surface spin with weaker AFM exchange resists flipping against the field until a lower field strength. On the other hand, the field at which the mod. surf $\downarrow$ monodomain begins to transition to a BL $\downarrow$ state for a ramping-down protocol is $-3.8~\mathrm{T}$, unchanged from the corresponding field for $x=-0.4$ in the main text. This is because this transition requires flipping the bottom spin of the slab to align with the negative field, which for our model of surface asymmetry always has an AFM exchange fixed to the bulk $J^{\mathrm{AFM}}$.\\
\indent The results for $x=0.8$ differ more dramatically from the results in the main text. First, we note that as expected we do get the opposite preferred domain for $x>0$ compared to $x<0$ for a fixed ramping direction. However, there is more oscillation in $\langle L_z\rangle$ as the system transitions from the spin-flop phase to the BL phase to an AFM monodomain phase, because unlike in the $x<0$ case, the bilayers locate in different positions across the five MC ensembles we use for averaging. Thus, the probability to reach the preferred mod. surf $\uparrow$ (for ramping-down) or mod. surf $\downarrow$ (for ramping-up), and thus the critical field at which it occurs, differs across the simulations in the ensemble. On the other hand, the critical field $H_z^*$ at which the AFM monodomain begins to transition to a BL state with surface spins in the opposite direction is quite consistent at $H^*\sim\pm 4.5\mathrm{T}$. This is because the field value of this second transition involves flipping of the bottom surface spin to align with the field and is set simply by the relative strengths of $J^{\mathrm{AFM}}$ and $H_z$ rather than a probabilistic motion and ``expelling" of the BL in the BL $\rightarrow$ AFM monodomain transition.\\
\indent Perhaps most notably, we see that for $x=0.8$, $\langle L_z\rangle$ at $H_z=0$ only reaches about $0.6$ in magnitude; this means that in our ensemble averages the AFM domain \textit{opposite} to that expected by our statistical mechanics arguments is selected about 2/5 of the time. This implies, in agreement with our scaling arguments in the main text and SI Sec.~E, that for such a thick slab of MBT, completely deterministic domain control via field ramping cannot be achieved when the surface exchange is more strongly AFM than the bulk $J^{\mathrm{AFM}}$. We note however that this vanishing of deterministic domain selection for thick slabs when $x>0$ assumes the other surface exchange retains the bulk coupling value, or also becomes inequivalently strengthened. Instead, if we look at the committor plot Fig.~\ref{fig:comm_bothsurf}(a) in SI Sec.~H, we see that the robustness for deterministic domain selection with a given surface having stronger AFM exchange than $J^{\mathrm{AFM}}$ becomes much stronger if the opposite surface has slightly \textit{weaker} AFM exchange, which is plausible for asymmetric growth conditions and environments, or defects.
\subsection{Monte Carlo simulations for six-layer $\mathrm{MnBi_2Te_4}$ slab }
 For our MC simulations of MBT in the main text, we used as thick a slab as possible to mimic the bulk-like sample in Ref.~\cite{Sunko2026}. In Fig. \ref{fig:MC_sixlayers} we compute the normalized ensemble-averaged N\'{e}el vector for a six-layer slab of MBT, corresponding to a thin film of about $8\mathrm{nm}$. The most obvious difference between results for the bulk-like 48-layer slab in the main text (Figs. 3 and 4) and SI Fig.~\ref{fig:MC_otherx}, and the thin film in Fig. \ref{fig:MC_sixlayers} is that we get deterministic domain selection ($\langle L_z\rangle=\pm1$) based on field-ramping direction at much smaller levels of surface anisotropy $x$ for a six-layer slab than the 48-layer slab. Specifically, we have consistent selection of opposite domains for a fixed ramping direction for $x=-0.1$ and $x=0.1$ respectively. In fact, additional MC tests which we do not show here indicated that for $x<0$, fully deterministic selection persists down to $x=-0.02$.\\
 \indent As mentioned in the main text and in SI Sec.~E, we would naively expect only a finite domain bias, rather than fully deterministic selection, for $x=+0.1$ even for the thin-film six-layer slab, since the lower-energy surface-localized bilayer state is energetically degenerate with the central BL position that leads to random domain selection. The fact that our MC simulations show fully deterministic selection for small values of $x>0$ in the six-layer slab is likely because the field-pinning of the surface spins, which occurs even before the central layers exits the spin-flop phase, makes it kinetically easier for a BL to nucleate at the surface. However, since our focus in the main text is on bulk-like thicknesses, we only mention this here to highlight the possibility that based on our MC simulations in Fig.~\ref{fig:MC_sixlayers}, deterministic domain selection should be achievable for both directions of surface exchange modification when the number of energetically degenerate central layers is sufficiently small to not outweigh the kinetic bias of the surface layer.\\
\indent In stark contrast to the small anisotropy values needed in Fig. \ref{fig:MC_sixlayers}, for the 48-layer slab studied in the main text, we first get $\langle L_z\rangle=\pm1$ at $H_z=0$ for $x=-0.4$, and we never get fully deterministic selection for $x>0$. The latter is consistent with the $~1/N$ scaling of domain biasing for the $x>0$ case discussed in SI Sec.~E. For $x<0$, the larger exchange anisotropy magnitude required for deterministic selection in the 48-layer slab in Figs.~3-4 of the main text compared to the six-layer slab in Fig.~\ref{fig:MC_sixlayers} is consistent with the fact that thicker AFMs have more possible central bilayer positions which, while statistically supressed to occur due to their higher energy compared to the BL state localized at the surface with modified exchange, still have a nonzero probability. As a result, a larger surface exchange anisotropy $x$ is required for thicker slabs in order to overcome this suppression in deterministic selection by increasing the energy difference between central BL positions and the surface-localized bilayer state.\\
\indent Other than the difference in necessary values of surface exchange anisotropy, the evolution of magnetic order as a function of $H_z$ is qualitatively similar for thin (six-layer) and thick (48-layer) MBT slabs based on our MC simulations. For $x=0.1$ and ramping-down, the BL $\uparrow$ state
$\rightarrow$ AFM and AFM $\rightarrow$ BL $\downarrow$ state transitions occur at $3.6\mathrm{T}$ and $-3.8\mathrm{T}$ respectively in Fig. \ref{fig:MC_sixlayers} (for ramping-up, the field signs are just swapped). For $x=-0.1$ and ramping down, the BL $\uparrow$ state
$\rightarrow$ AFM and AFM $\rightarrow$ BL $\downarrow$ state transitions occur at $3.4\mathrm{T}$ and $-3.6\mathrm{T}$ respectively. In both cases the critical fields for BL $\rightarrow$ AFM and AFM $\rightarrow$ BL transitions are nearly symmetric about $H_z=0$, in contrast to the 48-layer slab at $x=-0.4$ where the two directions of the transitions were fairly asymmetric ($|2.6|\mathrm{T}$ for BL $\rightarrow$ AFM and $|3.8|\mathrm{T}$ for AFM $\rightarrow$ BL). This is simply because the surface exchange asymmetry needed for the thicker slab is much larger, and (as discussed in the previous section), since the two directions of the transition necessarily involve flipping opposite surface spins, the critical fields for BL $\rightarrow$ AFM and AFM $\rightarrow$ BL become less symmetric about zero as the surface exchange asymmetry $x$ increases.\\

\subsection{Energetics for bilayer, AFM and spin-flop states}
\begin{figure*}
    \centering
   \includegraphics{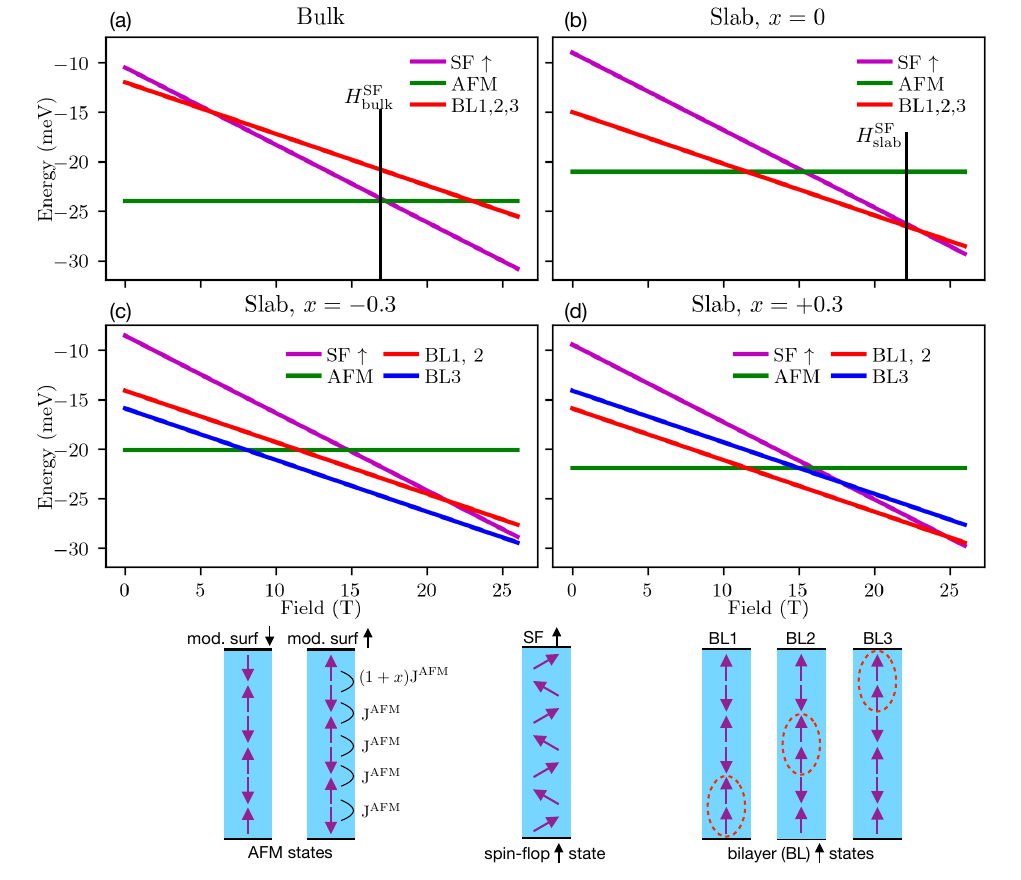}
   \caption{Energy of a six-layer slab from Eq.~\ref{eq:spinH_supp} versus field $H_z$ for the magnetic configurations in the bottom panel; note that the two AFM states, on the left of the bottom panel, are always energetically degenerate. Layers 1-5 have interlayer couplings $J^{\mathrm{AFM}}$, while the top coupling is $(1+x)J^{\mathrm{AFM}}$. (a) Periodic bulk AFM. (b) Symmetric vacuum-terminated slab ($x=0$). (c) Asymmetric slab with $x=-0.3$ ($J_{\mathrm{top}}=0.7J^{\mathrm{AFM}}$). (d) Asymmetric slab with $x=+0.3$ ($J_{\mathrm{top}}=1.3J^{\mathrm{AFM}}$).}
   \label{fig:bl_E}
\end{figure*}
\begin{figure}
    \vspace{0.5cm}
    \includegraphics{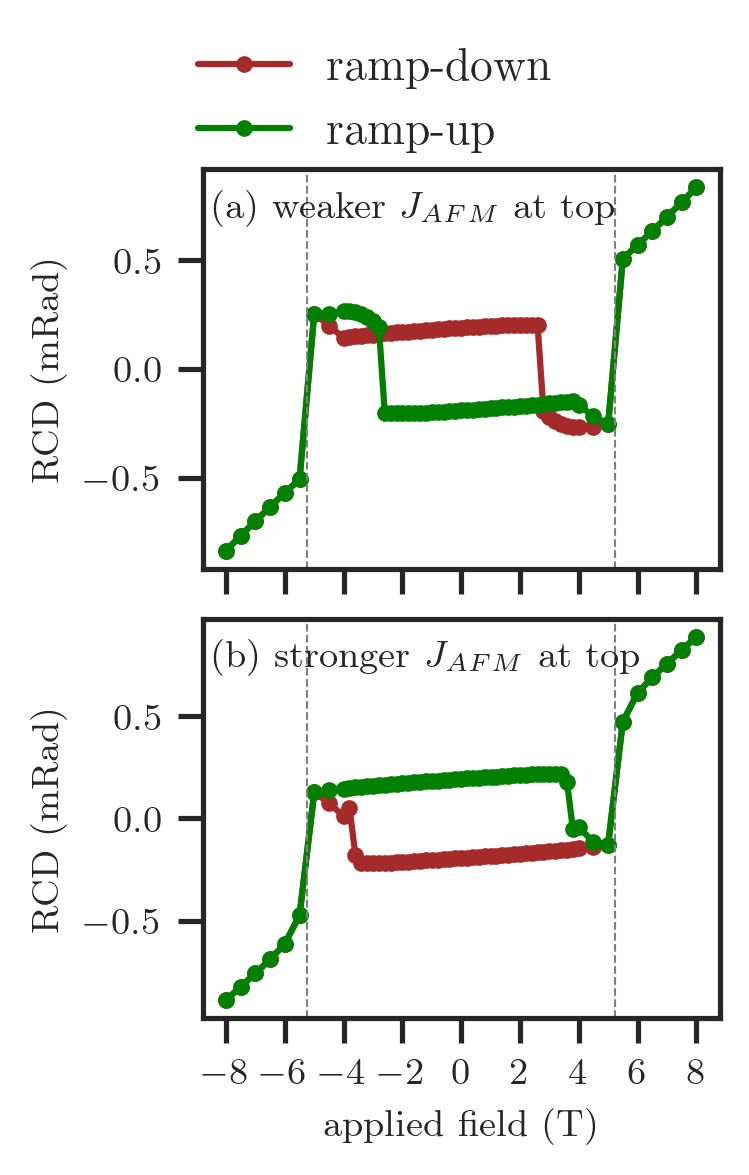}
    \caption{Simulated RCD measurements at 2.3eV for the MC magnetic structures found for $x=-0.4$ in the 48 layer slab. We show the results for the surface with (a) the weaker exchange and (b) the stronger exchange at the top of the slab (exposed to light). The dashed lines denote the spin-flop transition field, $H_z^{SF}\approx5.25\text{T}$.}
    \label{fig:RCD_SM}
\end{figure}

In the main text, we argued that for realistic AFM crystals with vacuum-terminated surfaces, spins at the surface have a reduced AFM exchange compared to sublattices in the bulk due to truncation of nearest neighbors. This implies that there is a range of fields for which a multi-domain bilayer (BL) state characterized by top and bottom surface spins pinned parallel to the direction of an applied field $H_z$ and a ferromagnetically coupled bilayer (equivalently, a domain wall) somewhere in the slab is energetically preferable to both the spin-flop phase and the monodomain AFM phase. Here we show this explicitly by plotting the energy as a function of $H_z$ for different magnetic configurations in our toy-model Hamiltonian (Eq.~1 in the main text). For convenience, we repeat the Hamiltonian here, which is
\begin{equation}
    E=E_0+J^{\mathrm{AFM}}\sum_{ij}\hat{e}_i\cdot\hat{e}_j-K\sum_i(\hat{e}_i\cdot\hat{z})^2-C\sum_i(S_i)^zH_z.
    \label{eq:spinH_supp}
\end{equation}
As in the main text, the sum is over magnetic atoms in the unit cell (with one atom per layer); $\hat{e}_i$ is the unit vector parallel to the magnetic moment on site $i$; $S_{i}$ is the magnetic moment of site $i$ in $\mu_B$, and $C=0.058\frac{\mathrm{meV}}{\mu_B\mathrm{T}}$ converts $\mu_B\cdot\mathrm{T}$ to eV.\\
\indent We take the same parameters which we used to evaluate the committors in SI Sec. E, namely $N=6$ for the number of AFM-coupled layers, $J^{\mathrm{AFM}}=3$ $\mathrm{meV}$, $|S|=4.5$ $\mu_B$, and $K=1$ $\mathrm{meV}$. In Fig.~\ref{fig:bl_E} we only consider positive values of $H_z$ (as we did in the cartoon Fig.~ 2 in the main text) which reflect a ramping-down protocol. Thus, the relevant states to consider, other than the two degenerate AFM domains (Fig.~\ref{fig:bl_E}, bottom left cartoons), are the ``spin-flop $\uparrow$" states (bottom center), where the N\'{e}el vector is perpendicular to $\hat{z}$ and the spins are canted toward $+\hat{z}$; and the BL $\uparrow$ states (bottom right) where top and bottom surface spins are pinned along $+\hat{z}$ and the ferromagnetically coupled bilayer has spins pointing up. We use the same labeling for the three possible BL states as in the main text; in particular, BL3 refers to the state where the bilayer is formed at the top surface which in our toy model is the surface which has an exchange differing from the bulk $J^{\mathrm{AFM}}$ (bottom left cartoon).\\
\indent First, in Fig.~\ref{fig:bl_E}(a) we plot the energy of these different magnetic configurations per six-layer unit cell as a function of applied field $H_z$ for a \textit{bulk} unit cell, where the top and bottom spins in the unit cell have periodic boundary conditions along $\hat{z}$. With our stated parameters we get a spin-flop field $H^{SF}_{\mathrm{bulk}}=17~\mathrm{T}$ at the point where the collinear AFM states become lower energy than the SF $\uparrow$ state. Note that to calculate the energy of a concrete spin-flop $\uparrow$ phase we have canted the spins at an arbitrary $45^{\circ}$ angle with respect to $\hat{z}$; thus, the exact $H^{SF}_{\mathrm{bulk}}$ in the model might shift up or down slightly depending on the actual lowest-energy angle of the spins in the spin-flop phase. Most importantly however, with periodic boundary conditions and no surface spins, there is no range of fields where the BL states are lowest in energy.\\
\indent In Fig.~\ref{fig:bl_E}(b), we plot the energy given by Eq. \ref{eq:spinH_supp} for the same six-layer unit cell, but this time we enforce vacuum boundary conditions such that the top and bottom spins in the unit cell each have only a single neighboring AFM-coupled layer. At this stage however, we assume the exchanges for top and bottom surface spins both remain equal to $J^{\mathrm{AFM}}$, corresponding to the surface exchange asymmetry $x=0$. Now, there is a region of fields between $\sim 10~\mathrm{T}-22~\mathrm{T}$ where the BL states, rather than the spin-flop or AFM states, are the lowest in energy. The field where the spin-flop $\uparrow$ state becomes lower than the BL states gets moved to higher values compared to the bulk case, $H^{SF}_{\mathrm{slab}}\sim22~\mathrm{T}$. However, because the surface exchanges are identical, all three possible BL states are degenerate in energy. As we argued intuitively and showed via MC simulations and committor evaluations in the main text and SI Secs.~F-H respectively, this degeneracy means that subsequent evolution to an AFM state as $H_z$ is decreased to zero is not deterministic when $x=0$.\\
\indent In Figs. \ref{fig:bl_E}(c) and (d), we plot energy vs. $H_z$ for vacuum-terminated slabs which have different AFM exchanges; in particular, as shown in the bottom left cartoon, we vary interlayer exchange at the top surface by a surface exchange asymmetry $x$ such that $J^{\mathrm{AFM}}\rightarrow (1+x)J^{\mathrm{AFM}}$ (identically to the main text). We include results for both negative and positive values of $x$, specificaly $x=-0.3$ and $x=+0.3$ in Figs. \ref{fig:bl_E}(c) and (d) respectively. Just like for the symmetric slab in Fig. \ref{fig:bl_E} in Fig. \ref{fig:bl_E}(b), we get a finite window of fields where BL states are energetically preferred. Now however, the BL3 state becomes either lower (c) or higher (d) in energy than the other BL states BL1 and BL2. As we showed in this work, this breaking of degeneracy for intermediate multidomain BL states leads to an inequivalence in the probabilities to transition to the two AFM monodomain states. The domain which is preferentially selected for a given value of $x$ will be opposite compared to the ramping-down case for the ramping-up direction, where the BL states will have spins pinned along $-\hat{z}$.\\
\indent Finally, we note that in choosing the value of magnetic field $H_z$ at which to evaluate physically meaningful committors to transition from any of the BL states to either AFM domain, it is necessary to choose a field magnitude at which are AFM states are lower-energy than the BL states, such that these transitions are probable. Thus, we chose $+4\mathrm{T}$ ($-4\mathrm{T}$) for ramping-down (ramping-up) protocols, since from Fig. \ref{fig:bl_E} and analogous figures we generated from $x=-0.5$ to $x=0.5$, it is clear that at these fields the AFM states are always lower energy than the BL states for the range of surface exchange asymmetries we consider.\\

\subsection{Simulations of MOKE using magnetization-field data from Monte Carlo}

In Fig.~4(b) of the main text we show RCD (i.e. imaginary part of MOKE) signal calculated for magnetic configurations found by MC simulations. To calculate the complex MOKE signal we employ the transfer matrix approach, and use dielectric properties of MBT. Here we briefly recount the relevant methodological aspects, while more details can be found in Ref.~\cite{Sunko2026}.

Within this approach, each layer is considered as a material whose dielectric properties depend on the layer magnetization. Conceptually, the reflectivity  matrix of such a slab is found by solving boundary conditions at each interface between layers. In practice, this is done computationally following the transfer matrix approach as described in Ref.~\cite{hendriks_enhancing_2021}. As in Ref.~\cite{Sunko2026}, we take the dielectric tensor  of MBT to be equal to 

\begin{equation}
    \varepsilon_C =
    \begin{pmatrix}
        \varepsilon_+ & 0 \\
        0 & \varepsilon_-,
    \end{pmatrix}
\end{equation}
in the basis or circular polarization. The diagonal elements of the $k$-th layer are given by the Lorentzian: 

\begin{equation}\label{eq:Lorentz}
\varepsilon^k_{\pm}(\omega) = \varepsilon_\infty + \frac{f}{(\omega_0 \pm\delta\omega\frac{m_z^k}{m^s})^2 - \omega^2 - i\gamma \omega},
\end{equation}

\noindent where $f$, $\omega_0$, and $\gamma$ are oscillator strength, resonant frequency, and damping. These parameters, together with a background contribution, $\varepsilon_\infty$, are chosen to reproduce the measured optical conductivity in the spectral range of interest, reported in Ref.~\cite{kopf_influence_2020}. The difference in resonance frequency for LC and RC polarized light for a saturated layer magnetization, $\delta\omega$, is chosen to fit the magnitude of the RCD spectrum at the spin-flop transition. Saturation magnetization is taken to be  $m^s=4.5\mu_B$, as obtained by DFT for the magnetic ground state~\cite{Sunko2026}, while $m_z^k$ is the average $z$ component of the Mn moment in layer $k$. The effective splitting between the resonant frequency for the RC and LC light for layer $k$ is then reduced from $\delta\omega$ by the ratio $m_z^k/m^s$. The MC simulation yields $m_z^k$ for each layer; this is how we calculate MOKE for the MC structures. All the dielectric parameters are kept the same as in Ref.~\cite{Sunko2026}; we reproduce them in Table~\ref{tab:optical_params} for convenience.

\begin{table}[h]
    \centering
      \renewcommand{\arraystretch}{1.2} 
    \setlength{\tabcolsep}{4pt} 
    \begin{tabular}{| c|  c|  c|  c | c|}
        \hline
         $\gamma / \omega_0$ & $ f / (\omega_0^2\varepsilon)$ & $\omega_0 [eV]$ & $\delta \omega [eV]$ & $\varepsilon_{\infty}/\varepsilon$ \\
        \hline
     
         0.424 & 4.2 & 2.05 & -0.0708 & 8.25 + 10.9 $i$ \\
    
        \hline
    \end{tabular}
    \caption{Parameters of the Lorentz model (Eq.~\ref{eq:Lorentz}), used for MOKE simulations. 
    }
    \label{tab:optical_params}
\end{table}

One more parameter needed for the MOKE calculation is the layer thickness. In Ref.~\cite{Sunko2026} the experimental value of $d_{exp}=1.3\text{nm}$ was used, and with realistic dielectric parameters (Tab.~\ref{tab:optical_params}) we found that the reflectance becomes insensitive to the number of layers for slabs thicker than $\approx 60$ layers, or $80\text{nm}$. The experimental slab thickness of $\approx130\text{nm}$ is therefore firmly in the bulk limit, while even the 48 layer thick MC slab is not. To compensate for this difference, and enable the comparison to the experiment, we set $d=2d_{exp}$ in the simulations - this ensures that the simulated MOKE is not sensitive to the bottom surface. Based on the understanding of MOKE in AFMs~\cite{Sunko2026}, we expect this quantitative change does not alter the qualitative picture.

We now have the tools to calculate MOKE for magnetic structures found by MC simulations, at any photon energy. First, we do this for the 48 layer slab with $x=-0.4$ and the photon energy of $2.3\text{eV}$, comparable to the value used for experiments reported in Fig.~2 of Ref.~\cite{Sunko2026}. In Figs.~\ref{fig:RCD_SM}(a,b) we plot RCD for two configurations: for light shining on the surface with weaker and stronger exchange, respectively. In both cases the sign of RCD is trainable by the field protocol, and the spin-flop transition is found at $H_z^{SF}\approx5.25\text{T}$, marked by the dashed lines. Further, in both cases the RCD exhibits discontinuities at fields of smaller magnitude than $H_z^{SF}$, denoted $H_z^*$ in the main text. The value of $H_z^*$, as well as the switching process itself, vary between the panels, reflecting different properties of probed surfaces.

We note that the simulation for more strongly coupled spins at the surface exposed to light (Fig.~\ref{fig:RCD_SM}(b)) agrees better with the experimental data in Ref.~\cite{Sunko2026}. This is consistent with the conclusion reached based on the spectroscopic study in Ref.~\cite{Sunko2026}, namely, that the surface spin exposed to light retains the orientation set by the field when the field is ramped towards $H_z=0$. In contrast, as we discuss in the main text, in the transition to the AFM monodomain the spins on the surface with weaker exchange will flip opposite to the direction of the field as it is ramped towards $H_z=0$.
\subsection{Comparing deterministic domain selection in compensated even-layer and uncompensated, odd-layer $\mathrm{MnBi_2Te_4}$ }
In the main text, we have emphasized how our proposed mechanism for deterministic magnetic field-based AFM domain selection in compensated AFMs with no magnetic moment is different from well-known AFM control methods based purely on exchange bias (EB) from uncompensated magnetization. Specifically, whereas EB involves breaking the energetic degeneracy between the two AFM monodomains due to interfacial coupling to a ferromagnetic layer, in our model the two AFM monodomains remain completely energetically degenerate despite the inequivalence in exchange between top and bottom surfaces. The analogy to exchange bias is limited to the intermediate step in which field-pinned surface spins favor a particular bilayer configuration. The final deterministic selection in the compensated slabs arises only subsequently, from biased transition pathways out of this intermediate state, while the two AFM monodomains remain energetically degenerate.\\
\indent Here, we concretely demonstrate with Monte Carlo (MC) simulations how our proposed domain selection mechanism, relevant for completely compensated AFMs with no net magnetization, differs from well-known EB mechanisms requiring a net magnetic moment in the overall system. Specifically, we compare results of field-ramping simulations of even-layer compensated MBT (the focus of this work) with those for \textit{ferrimagnetic} odd-layer MBT, which in its ground state has a net magnetization due to the odd number of AFM coupled layers. In both cases, we make the surface exchanges inequivalent as required in the mechanisms of focus in this work. This comparison between even- and odd-layer MBT is helpful, because it is the odd-layer MBT that actually bears strong resemblance to the `textbook' EB mechanism: it can be thought of as an even-layer compensated AFM, interfaced with a surface ferromagnetic layer. Indeed, odd-layer MBT does exhibit classic signatures of EB (see Ref.~\cite{Chong2024}).\\
\begin{figure*}
    \centering
   \includegraphics{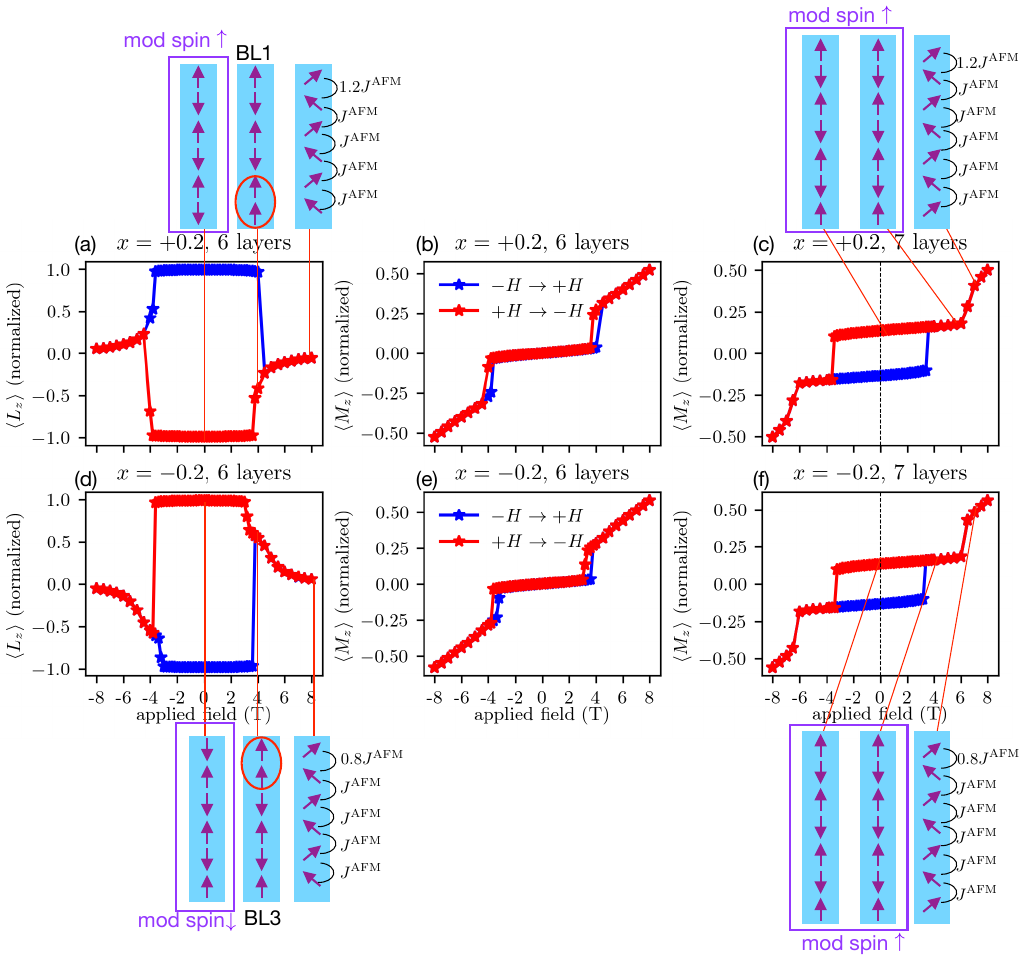}
    \caption{Monte Carlo results for magnetic field-based domain selection for six-layer MBT with no net magnetization versus ferrimagnetic seven-layer MBT with a net magnetic moment. In each panel, red curve represents ramping-down simulation, and blue represents ramping up. (a) Normalized, ensemble-averaged Néel vector $\langle L_z\rangle$ for six-layer MBT with surface exchange anisotropy $x=+0.2$ for the top surface exchange. (b) Normalized ensemble-averaged $\hat{z}$ component of net magnetization $\langle M_z\rangle$ for six-layer MBT with $x=+0.2$. (c) Normalized ensemble-averaged $\hat{z}$ component of net magnetization $\langle M_z\rangle$ for uncompensated, seven-layer MBT with $x=+0.2$. Figs. (d)-(f) are the same as (a)-(c) respectively, except all simulations use a surface exchange anisotropy of $x=-0.2$ for the topmost interlayer exchange. Cartoons show the layer-resolved magnetic states for the ramping-down protocol at different fields corresponding to different $\langle L_z\rangle$ and $\langle M_z\rangle$ values for six-layer and seven-layer systems respectively.}
    \label{fig:EB_netM}
\end{figure*}
\indent We perform MC simulations of ramping-down (red) and ramping-up (blue) field protocols for even-layer MBT composed of six septuple layers, and for odd-layer MBT composed of seven septuple layers, with all other magnetic and simulation parameters held the same as those used in the main text. In each case, we impose an inequivalence between the top- and bottom-most interlayer AFM exchanges, with exchanges $1$-$4$ for six-layer MBT and $1$-$5$ for seven layer MBT set equal to the bulk $J^{\mathrm{AFM}}$ and the remaining interlayer exchange set to $(1+x)J^{\mathrm{AFM}}$, as in the main text. In the first row of Fig.~\ref{fig:EB_netM}, $x$ is set to $x=+0.2$ and in the second row $x$ is set to $x=-0.2$. Figs. ~\ref{fig:EB_netM}(a), (b), (d) and (e) show results for compensated six-layer MBT, with (a),(d) showing the ensemble-averaged, normalized N\'{e}el vector $\langle L_z\rangle$, and (b),(e) the ensemble-averaged, normalized $\hat{z}$-component of layer magnetization $\langle M_z\rangle$. Figs.~\ref{fig:EB_netM}(c) and (f) show $\langle M_z\rangle$ for the uncompensated, seven-layer MBT slab. The cartoons for (a), (d) and (c),(f) show layer-resolved magnetization at varying field values in the ramp-down protocols, at field values indicated by red lines.\\
\indent There are several key points to mention. First, the favored monodomain is achieved immediately below $H_z^{\mathrm{SF}}$ for $N=7$, but not for $N=6$. Focusing on the ramping-down protocols (red), and the corresponding cartoons, we see that $\langle M_z\rangle $ is constant for $N=7$ between $0\leq H_z<H^{\mathrm{SF}}$ (panels (c) and (f); there is a small slope due to the finite susceptibility at non-zero temperatures). Both surface spins are parallel to the field direction at $H_z=0$, and a ferrimagnetic monodomain is selected. For $N=6$, our MC simulations show that the top and bottom surface layers are also pinned along $+z$ for the ramp-down protocol just below $H_z^{\mathrm{SF}}$, but this does not correspond to an AFM monodomain. This is clear both from the cartoons and, which show ferromagnetic bilayers, and $\langle L_z \rangle$, which does not saturate ((a) and (d)).\\
\indent A second key difference between our proposed method and EB is that our model does not require a weak uncompensated moment. This can be seen two different ways in Fig.~\ref{fig:EB_netM}. First, Figs.~\ref{fig:EB_netM}(b) and (e) show vanishing net magnetization at $H_z=0$, and no hysteresis around $H_z=0$, for both signs of the surface exchange anisotropy. Further, for $N=7$ the same monodomain is selected for both signs of the modified surface anisotropy (Figs.~\ref{fig:EB_netM}(c) and (f)). In contrast, for $N=6$ the monodomain selection depends on the sign of surface anisotropy (Figs.~\ref{fig:EB_netM}(a) and (d)). This is because for $N=7$ the monodomain selection proceeds due to field pinning of the uncompensated surface layer, set only by the initial field direction, whereas for $N=6$ there is no net moment in the AFM monodomain state, and the modified surface exchange is essential. Even if there was a small uncompensated moment in the $N=6$ slab, the fact that the domain choice depends on the surface anisotropy means that this small moment is not driving the domain selection.


\end{document}